\documentclass[aps,floats,prd,twocolumn,nofootinbib]{revtex4-2}

\usepackage{amssymb}
\usepackage{amsmath}

\usepackage{graphicx}
\usepackage{graphics}
\usepackage[caption=false]{subfig}
\usepackage{dcolumn}
\usepackage{color}
\usepackage{fancyhdr} 
\usepackage{hyperref}

\usepackage[utf8]{inputenc}
\DeclareUnicodeCharacter{00A0}{ }

\makeatletter
\renewcommand*{\fnum@figure}{{\normalfont\bfseries \figurename~\thefigure}}
\renewcommand*{\@caption@fignum@sep}{\textbf{ : }}
\makeatother

\makeatletter
\renewcommand*{\fnum@table}{{\normalfont\bfseries \tablename~\thetable}}
\makeatother

\def\VEV#1{\left\langle #1 \right\rangle}

    \newcommand{\be}{\begin{equation}}
  \newcommand{\ee}{\end{equation}}
    \newcommand{\ba}{\begin{align}}
  \newcommand{\ea}{\end{align}}

\newcommand{\sinc}{\mathrm{sinc}}

\newcommand{\vcb}{ v_{\rm cb}  }

\begin{document}

\title{Robust Velocity-induced Acoustic Oscillations at Cosmic Dawn}

\author{Julian B.~Mu\~noz\footnote{Electronic address: \tt julianmunoz@fas.harvard.edu}
} 
\affiliation{Department of Physics, Harvard University, 17 Oxford St., Cambridge, MA 02138}

\date{\today}

\begin{abstract}
The redshifted 21-cm line of hydrogen holds great potential for the study of cosmology, as it can probe otherwise unobservable cosmic epochs.
In particular, 	measurements of the 21-cm power spectrum during cosmic dawn---the era when the first stars were formed---will provide us with a wealth of information about the astrophysics of stellar formation, 
as well as the origin of fluctuations in our Universe.
In addition to their usually considered density fluctuations, dark matter and baryons possess large relative velocities with respect to each other, due to the baryon acoustic oscillations (BAOs) suffered by the latter, which suppress the formation of stars during cosmic dawn, leaving an imprint on 21-cm observables during this era.
Here we present {\tt 21cmvFAST}, a version of the publicly available code {\tt 21cmFAST} modified to account for this effect. 
Previous work has shown that the inclusion of relative velocities produces an acoustic modulation on the large-scale 21-cm power spectrum during cosmic dawn.
By comparing analytic calculations with simulations from {\tt 21cmvFAST}, here we demonstrate that this modulation takes the form of robust velocity-induced acoustic oscillations (VAOs), during both the Lyman-$\alpha$ coupling era and the subsequent epoch of heating.
The unique shape of these VAOs, which is determined by the acoustic physics that generated the relative velocities, can be analytically computed and is easily distinguishable from the usual density-sourced fluctuations. 
We find that, under a broad range of astrophysical assumptions, the VAOs are detectable at high significance by the upcoming HERA interferometer, which could therefore confirm the presence of acoustic oscillations at cosmic dawn.
\end{abstract}

\maketitle

\section{Introduction}

The matter in our Universe clusters under its own gravitational pull, slowly growing the seeds of primordial fluctuations into the rich large-scale structure of our cosmos~\cite{Bardeen:1983qw,Mukhanov:1987ny,Ma:1995ey,Springel:2005nw}.
Nonetheless,  not all matter behaves equally. 
While the majority of matter  is dark, cold, and collisionless, roughly a sixth of it is baryons, capable of interacting with photons through electromagnetic forces.
These interactions give rise to the famous baryon acoustic oscillations (BAOs), which are imprinted onto the distribution of baryons thereafter and have been observed in the cosmic microwave background (CMB)~\cite{Aghanim:2018eyx}, as well as at low-redshift surveys~\cite{Aaboud:2016tnv,Alam:2016hwk,Ata:2017dya}.

Interestingly, the lack of BAOs for the dark matter (DM) gives rise to a bulk relative velocity between it and the baryons~\cite{Tseliakhovich:2010bj}.
This ``streaming" velocity, with a root-mean-squared value of $v_{\rm rms}\approx 30$ km s$^{-1}$ at recombination, becomes supersonic thereafter due to the low baryonic temperature.
The resulting supersonic motions between baryons and DM---which fluctuate on acoustic scales---have three suppressive effects on the formation of the first stars, as $(i)$ fewer haloes are formed~\cite{Tseliakhovich:2010bj,Naoz:2011if,Bovy:2012af}, $(ii)$ the amount of gas in each halo available for cooling is smaller~\cite{Tseliakhovich:2010yw,Dalal:2010yt,Naoz:2012fr}, and $(iii)$ it is more difficult for said gas to settle down into stars~\cite{Greif:2011iv,Stacy:2010gg,OLeary:2012gem,Fialkov:2011iw,Schauer:2018iig,Hirano:2017znw}.
Together, these effects modulate the number of star-forming galaxies in the early universe as a function of the DM-baryon relative velocity~\cite{Barkana:2016nyr,Fialkov:2012su,Fialkov:2013jxh,McQuinn:2012rt,Visbal:2012aw}.

Even though we do not have direct access to the first star-forming galaxies at redshift $z \sim 20$, the hydrogen 21-cm line will allow us to indirectly map the distribution of the first stars, as their emission affects the 21-cm brightness temperature~\cite{Furlanetto:2006jb,Pritchard:2011xb,Mirocha:2017xxz}.
The presence of streaming velocities is often ignored in public 21-cm codes, such as {\tt 21cmFAST}~\cite{Mesinger:2010ne,Greig:2015qca}, which is a good approximation if star formation mainly occurs in fairly massive haloes (with virial temperatures $T_{\rm cool} \gtrsim {\rm few} \times 10^4$ K), but will cause a severe overestimation if there is stellar formation in molecular-cooling haloes (with $T_{\rm cool} \lesssim 10^4$ K~\cite{Abel:2001pr,Bromm:2003vv,Haiman:2006si}).
Here we present {\tt 21cmvFAST}\footnote{\url{ https://github.com/JulianBMunoz/21cmvFAST} }, a modified version of {\tt 21cmFAST}\footnote{\url{ https://github.com/andreimesinger/21cmFAST} } which properly accounts for the effects of the DM-baryon relative velocities.

Building upon the work of Refs.~\cite{Tseliakhovich:2010bj,Dalal:2010yt,McQuinn:2012rt,Visbal:2012aw}, we include relative velocities in the initial conditions of {\tt 21cmvFAST}, which then affects the amount of stellar formation at every cell in the simulation box. 
We additionally account for the averaged effect of Lyman-Werner (LW) feedback, which dissociates molecular hydrogen and raises the cooling temperature of haloes, reducing the effect of the streaming velocities at lower redshifts~\cite{Machacek:2000us,Fialkov:2012su,Visbal:2014fta}. 
Thus, while for studies of the epoch of reionization, at $z\sim 6$, the streaming velocities are fairly  unimportant (see, however, Ref.~\cite{Cohen:2015qta}), they have a large impact on the 21-cm signal during the cosmic-dawn era ($z = 15-30$),
as shown in Refs.~\cite{Dalal:2010yt,McQuinn:2012rt,Visbal:2012aw,McQuinn:2012rt,Fialkov:2012su,Fialkov:2013jxh}
and as we will further explore here.

We find that the DM-baryon relative velocities ($\vcb$) produce a modest delay $\Delta z \sim 2$ on all the landmarks of cosmic evolution, as fewer stars form on average.
More interestingly, however, the relative velocities show acoustic oscillations in their power spectrum, due to their BAO origin.
These oscillations are imprinted in the distribution of the first stars, and therefore in the 21-cm power spectrum during cosmic dawn.
This was first proposed in Ref.~\cite{Dalal:2010yt} during the Lyman-$\alpha$ coupling era (LCE), where the Lyman-$\alpha$ photons emitted by the first stars coupled the thermal and spin temperatures of neutral hydrogen~\cite{Wout,Field,Hirata:2005mz}.
A similar prediction was made in Refs.~\cite{Visbal:2012aw,McQuinn:2012rt} for the epoch of heating (EoH), where the X-rays emitted by the first galaxies heated the neutral hydrogen, thus affecting the 21-cm signal.
Here we use {\tt 21cmvFAST} to study the effect of the relative velocities across the entire cosmic-dawn era ($z=15-30$), comparing our simulations with analytic calculations.

Given the large astrophysical uncertainties during this epoch, we consider a broad range of possible feedback strengths. 
In all cases we find velocity-induced acoustic oscillations (VAOs) on the 21-cm power spectrum, with an amplitude growing in size during the LCE (peaking at $z\approx 24$ for our fiducial parameters), vanishing in the transition to the EoH, then peaking again during the heating era (at $z\approx 17$), and vanishing after the intergalactic medium (IGM) is fully heated ($z\lesssim 12$). 
This is in agreement with our analytic estimates.
Furthermore, we confirm that the shape of the VAOs follows the well-understood fluctuations in $\vcb^2$~\cite{Dalal:2010yt,Ali-Haimoud:2013hpa,Munoz:2018jwq}, 
with damping at small scales due to the nonlocality of photon propagation.
We parametrize this effect through a window function, as in Ref.~\cite{Dalal:2010yt}, which dampens VAOs below $k\approx 0.05$ Mpc$^{-1}$ during the LCE, due to the large mean-free path of UV photons; and below $k\approx 0.2$ Mpc$^{-1}$ during the EoH, as X-ray photons typically get absorbed closer to their source~\cite{Furlanetto:2009uf}.
Our analytically calculated VAOs and window functions very well match our numerical simulations during both the LCE and the EoH.

Finally, we study the observability of the VAOs with the upcoming hydrogen epoch of reionization array (HERA)~\cite{DeBoer:2016tnn}.
We find that under most of our astrophysical scenarios, the VAOs are observable at large ($>5$) signal-to-noise ratios.
This would confirm that the ``small" molecular-cooling galaxies are driving stellar formation at cosmic dawn, and 
additionally yield the first measurement of baryon acoustic oscillations at a redshift intermediate between galaxy surveys and the CMB,
which in Ref.~\cite{PaperII} we propose as a novel standard ruler during cosmic dawn.

This paper is structured as follows. In Section~\ref{sec:vcbeffect} we review the effects of the streaming velocities on the formation of the first stars.
We use these effects in Section~\ref{sec:21cmline} to find the changes in the 21-cm signal due to velocities, using {\tt 21cmvFAST} to find 21-cm maps and power spectra.
We quantify the effect of the velocities in detail in Section~\ref{sec:quantifyvcb}, and study its observability in Section~\ref{sec:obs}.
We conclude in Section~\ref{sec:conclusions}.

\section{The effect of velocities on the first stars}
\label{sec:vcbeffect}

We begin by reviewing the effect of the streaming velocities on the first star-forming galaxies.
This section heavily draws from Refs.~\cite{Tseliakhovich:2010bj,Tseliakhovich:2010yw,Dalal:2010yt,Fialkov:2011iw}, so the reader might want to skip to Section~\ref{sec:21cmline} to see the impact on the observable 21-cm line.
Throughout this work we use the best-fit cosmological parameters from Planck 2018~\cite{Aghanim:2018eyx}, and all lengths and wavenumbers will be comoving unless otherwise specified.

As in Ref.~\cite{Tseliakhovich:2010bj}, we use a moving-mesh perturbation theory, which is valid for regions of radius $R\lesssim 3$ Mpc, where the velocity is coherent.
Then, there exists a zero-order solution to the Boltzmann equations, with no overdensities,
\be
\delta_c = \delta_b  = 0 
\ee
but a nonzero relative velocity between the DM and baryon fluids
\be
\mathbf \vcb (z) =  \mathbf v^i_{\rm cb} \dfrac{1+z}{1+z_{\rm kin}},
\ee
where $\mathbf v^i_{\rm cb} $ is the value of the DM-baryon relative velocity at kinematic decoupling, $z_{\rm kin}$~\cite{Hirata:2017ivs}.
Additionally, we are free to choose a reference frame, which we set at the DM rest frame by setting
\begin{subequations}
\ba
\mathbf v_b &= - \mathbf \vcb (z) \\
\mathbf v_c &= 0,
\end{align}
\end{subequations}
which allows for easier computation at high wavenumbers $k$, where the baryon fluctuations oscillate rapidly~\cite{McQuinn:2012rt}.
Then, with this background solution we can rewrite the first-order Boltzmann equations from Ref.~\cite{Tseliakhovich:2010bj} as
\ba
\delta_c' &= \dfrac{\theta_c}{H(1+z)}  \nonumber \\ 
\theta_c' &= \dfrac{2 \theta_c}{(1+z)} + \dfrac{3 H}{2(1+z)} \,\delta_m   \nonumber \\
\delta_b' &= \dfrac{\theta_b}{H(1+z)}  + \dfrac{i \vcb k \mu}{H} \delta_b \label{eq:ODEs} \\
\theta_b' &= \dfrac{2 \theta_b}{(1+z)} + \dfrac{3 H}{2(1+z)} \,\delta_m  - \dfrac{c_s^2 k^2 (1+z)}{H} \left(\delta_b + \delta_T \right) + \dfrac{i \vcb k \mu}{H} \theta_b \nonumber \\
\delta_T' &= \dfrac{2}{3} \delta_b' + \dfrac{T_\gamma}{T_g} \dfrac{\Gamma_C}{H(1+z)} \delta_T, \nonumber
\end{align}
where $\theta_i$ is the velocity divergence of species $i$, $\mu \equiv \mathbf k \cdot \mathbf \vcb/(k\vcb)$, $H$ is the Hubble expansion rate at redshift $z$, $\vcb = |\mathbf \vcb|$, and the prime denotes derivative with respect to redshift.
Additionally, $\Gamma_C$ is the Compton interaction rate, $T_\gamma$ is the CMB temperature, and we include the effect of temperature fluctuations $\delta_T = \delta T_g/T_g$, where $T_g$ is the gas temperature. 
Here, the isothermal sound speed is given by~\cite{Ali-Haimoud:2013hpa}
\be
c_s^2 = \dfrac{k_B\,T_g}{\mu_b},
\ee
where $\mu_b$ is the average molecular mass of baryons and $k_B$ is the Boltzmann constant.

We solve Eqs.~\eqref{eq:ODEs} for different values of $v_{\rm cb}^i$ and $\mu$, by setting the initial conditions to be $\delta_b = \mathcal T_b$, $\delta_c= \mathcal T_c$, and $\delta_T=0$, where $\mathcal T_X$ is the transfer function, which evolves the initial primordial fluctuations to over/underdensities in the fluid $X$, obtained with the Boltzmann solver {\tt CLASS}~\cite{Blas:2011rf}. With this we obtain the baryon and DM transfer functions $\mathcal T_b(k,z,v_{\rm cb}^i)$ and $\mathcal T_c(k,z,v_{\rm cb}^i)$, at arbitrary redshifts and relative velocities, which we will use to find the properties of the first star-forming haloes.
Additionally, we define the matter transfer function $\mathcal T_m$ as usual, though due to their relative velocities the baryon and DM parts will have different phases.

Regions with large DM-baryon relative velocity $\vcb$ form fewer stars through three mechanisms.
First, less gas is able to collapse into the haloes, as the relative velocity allows gas to stream away.
Second, the cooling of said gas in each halo is impeded by the relative velocity, raising the minimum halo mass necessary to form stars.
Third, larger velocities result in lower matter fluctuations, due to the additional nonthermal ``pressure", and thus fewer haloes are formed.
For a recent review of these effects see, for instance, Ref.~\cite{Barkana:2016nyr}.
We now compute each of these effects in turn.

\subsection{Gas Fraction}

As in Refs.~\cite{Tseliakhovich:2010yw,Dalal:2010yt}, we calculate the effect of the DM-baryon relative velocity on the fraction of gas accreted by DM haloes through the filter mass $M_F$.
This quantity, first introduced in Ref.~\cite{Gnedin:2000uj}, allows us to compute the effect of the baryonic pressure, be it thermal or not, on the collapse of baryons to haloes, by using results from the linear regime.
As in Refs.~\cite{Naoz:2006ye,Tseliakhovich:2010yw} we fit the ratio
\be
\dfrac{\delta_b}{ \delta_m}= 1 + r_{\rm LSS} - k^2/k_F^2
\ee
to first order in $k^2$, where we use the notation of Ref.~\cite{Naoz:2006ye} in defining $r_{\rm LSS}$ as the low-$k$  large-scale structure (LSS) difference between baryons and all matter (which, however, we take at $k=1$ Mpc$^{-1}$ to avoid the BAO wiggles). We define the filter mass as~\cite{Tseliakhovich:2010yw}
\be
M_F \equiv \dfrac{4 \pi}{3} \rho_m^0 r_F^{3},
\ee
where $\rho_m^0$ is the matter density today, and the (comoving) filter scale is given by $r_F = \pi/k_F$.

While this fit is not perfect, it provides us with a good estimate of $M_F$ by using linear theory alone.
Still, there are two important caveats to note here. 
The first is that the effect of the $\vcb$-induced ``pressure" in the baryon small-scale distribution has a different $k$-behavior from the usual thermal pressure, although we parametrize it in the same way.
The second is that the filter mass is a real-space quantity, whereas the overdensities $\delta_b$ and $\delta_m$ that we computed are Fourier-space quantities. 
Therefore, we have to average over the (cosine of the) angle $\mu$ between $\mathbf k$ and $\mathbf \vcb$ at some point, since real-space quantities cannot depend on it. 
We choose to average the ratio $\delta_b/\delta_m$ directly, instead of each density individually, although we have checked that this produces differences below 10\% on the filter mass.
Given the filter mass we can calculate the amount of matter in each halo that is composed of gas through the fit
\be
f_{\rm gas} (M,\vcb) = f_b^{(0)} \left[ 1 + (2^{a/3} - 1) \left(\dfrac{M_F(\vcb)}{M}\right)^a\right]^{-3/a},
\ee
where $a=0.7$, and $f_b^{(0)} = f_b (1+ 3.2 r_{\rm LSS}) $~\cite{Tseliakhovich:2010yw,Barkana:2000fd}.

Notice that, in addition to the effects calculated here, there can be resonant amplification at very high wavenumbers $k\sim10^3$ Mpc$^{-1}$, as proposed in Ref.~\cite{Ali-Haimoud:2013hpa}, which are only relevant at scales below the haloes that we are interested in and thus can safely be ignored.

\subsection{Minimum Halo Mass}

In addition to lowering the fraction of  baryonic gas in each halo, the relative velocity has a critical effect on the formation of stars in small-mass haloes through H$_2$ molecular cooling.
By softening the large gas overdensities that form in the core of these haloes, the relative velocities make it more difficult to form stars~\cite{Stacy:2010gg,Greif:2011iv,OLeary:2012gem,Fialkov:2011iw,Hirano:2017znw,Schauer:2018iig}.
We use the parametrization of Ref.~\cite{Fialkov:2011iw}, where this effect is cast as an increase in the minimum virial velocity that a halo requires to cool through molecular hydrogen as
\be
V_{\rm cool}(z,\vcb) = \left[ \left(V_{\rm cool}^{(0)}\right)^2 + \alpha^2 \vcb^2(z)\right]^{1/2},
\label{eq:vcool}
\ee
where we take the well-known ($\vcb=0$) result of $V_{\rm cool}^{(0)}=4.0$ km s$^{-1}$ for molecular cooling, as it lies in between the results of the different simulations from Refs.~\cite{Stacy:2010gg,Greif:2011iv,Machacek:2000us}.
The parameter $\alpha$ accounts for the effect of streaming velocities, and it is expected to lie in the range $\alpha=4-6$~\cite{OLeary:2012gem}.
We will study two different values within this range, with $\alpha=4$ as in Ref.~\cite{Fialkov:2011iw} (fitted from the simulations of Refs.~\cite{Stacy:2010gg,Greif:2011iv}), or $\alpha=6$ as recent simulations seem to indicate~\cite{Hirano:2017znw,Schauer:2018iig}.
From $V_{\rm cool}$ we can define a minimum halo mass $M_{\rm cool}$ that can form stars as~\cite{Barkana:2000fd}
\be
M_{\rm cool}(z,\vcb) = 7 \times 10^6 M_\odot \left(\dfrac{V_{\rm cool}(z,\vcb)}{10\,\rm km\,s^{-1}}\right)^3 \left(\dfrac{1+z}{20}\right)^{-3/2}, 
\ee
during matter domination, and assuming our fiducial cosmological parameters, as only haloes with $M>M_{\rm cool}(z,\vcb)$ have virial temperatures large enough to produce stellar formation.

We additionally include the effect of Lyman-Werner (LW) feedback in the formation of stars. 
As gas collapses into stars, these will emit UV photons in the Lyman-Werner band (11.2$-$13.6 eV), which can easily photodissociate H$_2$~\cite{Draine:1996hna,Haiman:1996rc,Ricotti:2000at,Yoshida:2003rw,Bromm:2003vv}.
This raises the minimum halo mass as the Universe evolves~\cite{Machacek:2000us,Wise:2007nb,OShea:2006eik,SafranekShrader:2012qn}, until it reaches the atomic-cooling threshold, which we take to be at $V^{\rm HI}_{\rm cool} = 17$ km s$^{-1}$~\cite{Oh:2001ex,Bromm:2002hb} (corresponding to $T_{\rm cool} =10^4$ K and $M_{\rm cool} \approx 3 \times 10^7\, M_\odot$).
The LW band overlaps with the Lyman-$\alpha$ band that produces Wouthuysen-Field coupling, which we discuss in Section~\ref{sec:21cmline}, albeit the relation between the two fluxes depends on astrophysical assumptions regarding the photon escape fraction, as well as self-shielding, among others~\cite{Ahn:2012sb,1701.07031}. 
We will not attempt to model this process, and instead rely on the results of Refs.~\cite{Machacek:2000us,Wise:2007nb,Visbal:2014fta}, 
where it was shown that the effect of LW feedback is to raise the mass of cooling haloes as
\be
M_{\rm cool} (z,\vcb,F_{\rm LW}) = M_{\rm cool} (z,\vcb,0) \times \left[1 + B\, (F_{\rm LW})^\beta \right],
\label{eq:McoolLW}
\ee
where $M_{\rm cool} (z,\vcb,0)$ is the result without the LW feedback calculated above, and $F_{\rm LW} \equiv 4\pi J_{\rm LW}$ is the LW flux, with $J_{\rm LW}$ in the customary units of $10^{-21}$ ergs s$^{-1}$ cm$^{-2}$ Hz$^{-1}$ sr$^{-1}$. 
Here, the two parameters $B$ and $\beta$ control the effect of the LW feedback on the cooling of gas in haloes.

Given the large astrophysical uncertainties in both the effect of velocities and the strength of the LW feedback, we will study four possible feedback strengths:

$\bullet$ No feedback, $B=0$. We do not expect this case to accurately reproduce our Universe, but it is a useful toy model where the effect of the velocities is most prominent in the 21-cm power spectrum.

$\bullet$ Low feedback, $B=4$ and $\beta=0.47$. This model follows the fits found in Refs.~\cite{Machacek:2000us,Wise:2007nb}, albeit with a coefficient $B$ roughly half of the best-fit value.

$\bullet$ Regular feedback, $B=7$ and $\beta=0.47$. This is the prescription in Refs.~\cite{Fialkov:2012su,Visbal:2014fta}, as a fit to the simulations of Refs.~\cite{Machacek:2000us,Wise:2007nb}, which results in a larger impact of $F_{\rm LW}$ on the minimum halo mass. 

$\bullet$ High feedback, $B=7$ and $\beta=0.47$ again; however, here we assume that the effect of LW photons and the relative velocities is fully uncorrelated, as in Ref.~\cite{McQuinn:2012rt}, so that $V_{\rm cool}^{(0)}  (F_{\rm LW}) = V_{\rm cool}^{(0)} (0) \times \left[1 + B\, (F_{\rm LW})^\beta \right]^{1/3}$ instead of the formulation of Eq.~\eqref{eq:McoolLW}. 
This dampens the effect of velocities considerably, as we will see.

In each of these models we will take the average LW flux from the strong-feedback case in Ref.~\cite{Fialkov:2012su}, where it is self-consistently computed through the amount of baryons collapsed into stars as a function of redshift, except for the low feedback strength, where we divide it by a factor of 2.
Additionally, we take $\alpha=4$ in the regular- and high-feedback cases, and $\alpha=6$ in the no- and low-feedback cases, which enhances the $\vcb$-induced fluctuations.
We leave for future work to self-consistently compute the Lyman-Werner feedback in each cell of our {\tt 21cmvFAST} simulation, similarly to how the Lyman-$\alpha$ flux is computed.

Note that we parametrize the unknown strength of the LW feedback through the poorly understood fitting function Eq.~\eqref{eq:McoolLW}  ($B$ and $\beta$), where the coefficients are fit from a few (three) datapoints in Refs.~\cite{Machacek:2000us,Wise:2007nb}, as opposed to the redshift at which $F_{\rm LW}$ is evaluated, as in Ref.~\cite{Fialkov:2012su}, since in Ref.~\cite{Visbal:2014fta} it was found that evaluating at the redshift of collapse is a good approximation unless $J_{21}$ grows extremely rapidly.

\subsubsection*{Correction to $f_*$}

Following Refs.~\cite{Machacek:2000us,Fialkov:2012su}, we assume that not all haloes above $M_{\rm cool}$ form stars with the same efficiency, as haloes near the threshold have a smaller fraction of their gas that can cool.
We follow the parametrization in Ref.~\cite{Fialkov:2012su}, where the fraction of baryons that form stars depends on the halo mass $M$ as
\be
f_*(M) = \alpha_* \log(M/M_{\rm cool})
\label{eq:fstar}
\ee
for $M_{\rm atom} > M > M_{\rm cool}$, and $f_* = f_*^0$ for $M>M_{\rm atom}$, where $f_*^0$ is the usual input fraction of baryons in stars in {\tt 21cmvFAST}.
We set the scale at which the fraction plateaus at the mass $M_{\rm atom}$ of haloes that can form stars through atomic cooling, as in Ref.~\cite{Fialkov:2012su}, which sets $\alpha_* = f_*^0/\log(M_{\rm atom}/M_{\rm cool})$ for $M_{\rm cool} < M_{\rm atom}$.
We will take $f^0_*=0.1$ throughout, except in the unrealistic no-feedback case, where we do not include this log-correction, which we compensate by taking a lower value of $f^0_*=0.03$.
For reference, $M_{\rm atom}\gg M_{\rm cool}$ at $z\geq 20$ for all feedback cases, whereas by $z=15$ the LW flux in the regular-feedback case is large enough to almost saturate, producing $M_{\rm cool} \approx M_{\rm atom}$.

\subsection{Halo Mass Function}

Through providing an additional source of pressure for baryons at small scales, the DM-baryon relative velocities can greatly affect the abundance---and bias---of the first star-forming haloes of our Universe, as first pointed out in Ref.~\cite{Tseliakhovich:2010bj}, and confirmed with simulations in Ref.~\cite{Naoz:2011if}.
We follow their calculation here, albeit using the Reed halo mass function (HMF) from Ref.~\cite{Reed:2006rw}, which is similar to the Sheth-Tormen HMF~\cite{Sheth:1999su}, but better calibrated at the redshifts of interest ($z\sim 10-30$) to correctly account for a suppression of high-mass haloes.

We begin by calculating the local value of the matter power spectrum~\cite{Tseliakhovich:2010bj}
\be
P_m^{\rm loc} (k,z,\vcb) = P_\zeta(k) \int_{-1}^1 \dfrac{d\mu}{2}  \left | \mathcal T_m(k,z,\vcb,\mu) \right|^2,
\label{eq:LocalPm}
\ee
by averaging over angles $\mu$, where $P_\zeta(k)$ is the primordial power spectrum.
Then, we define the variance in matter overdensities that will form haloes of mass $M$ as
\be
\sigma_M^2(z,\vcb) = \int \dfrac{d^3k}{(2\pi)^3} |W(k R_M)|^2 P_m^{\rm loc} (k,z,\vcb),
\ee
where $R_M = [\Omega_M H_0^2/(2M)]^{-1/3}$ is the typical radius of these overdensities (with $H_0$ the Hubble expansion rate today and $\Omega_M$ the cosmic matter abundance), which we assume have a top-hat shape, so the window function is given by $W(x)=3 x^{-2} [\cos(x)-\sin(x) x^{-1}]$.
Note that we find $P_m$ by solving Eqs.~\eqref{eq:ODEs} starting at the baryon drag era ($z_d=1060$), as suggested in Refs.~\cite{McQuinn:2012rt,Ahn:2018rkq}, as otherwise we would be underestimating the suppression due to velocities.
Nonetheless, as {\tt 21cmFAST} does not keep track of baryons and dark matter independently we cannot include any additional suppression due to differences in $\delta_b-\delta_c$~\cite{Schmidt:2016coo,Ahn:2016bcr}, which could further enhance the effects of the streaming velocities.

\begin{figure}[hbtp]
	\includegraphics[width=0.48\textwidth]{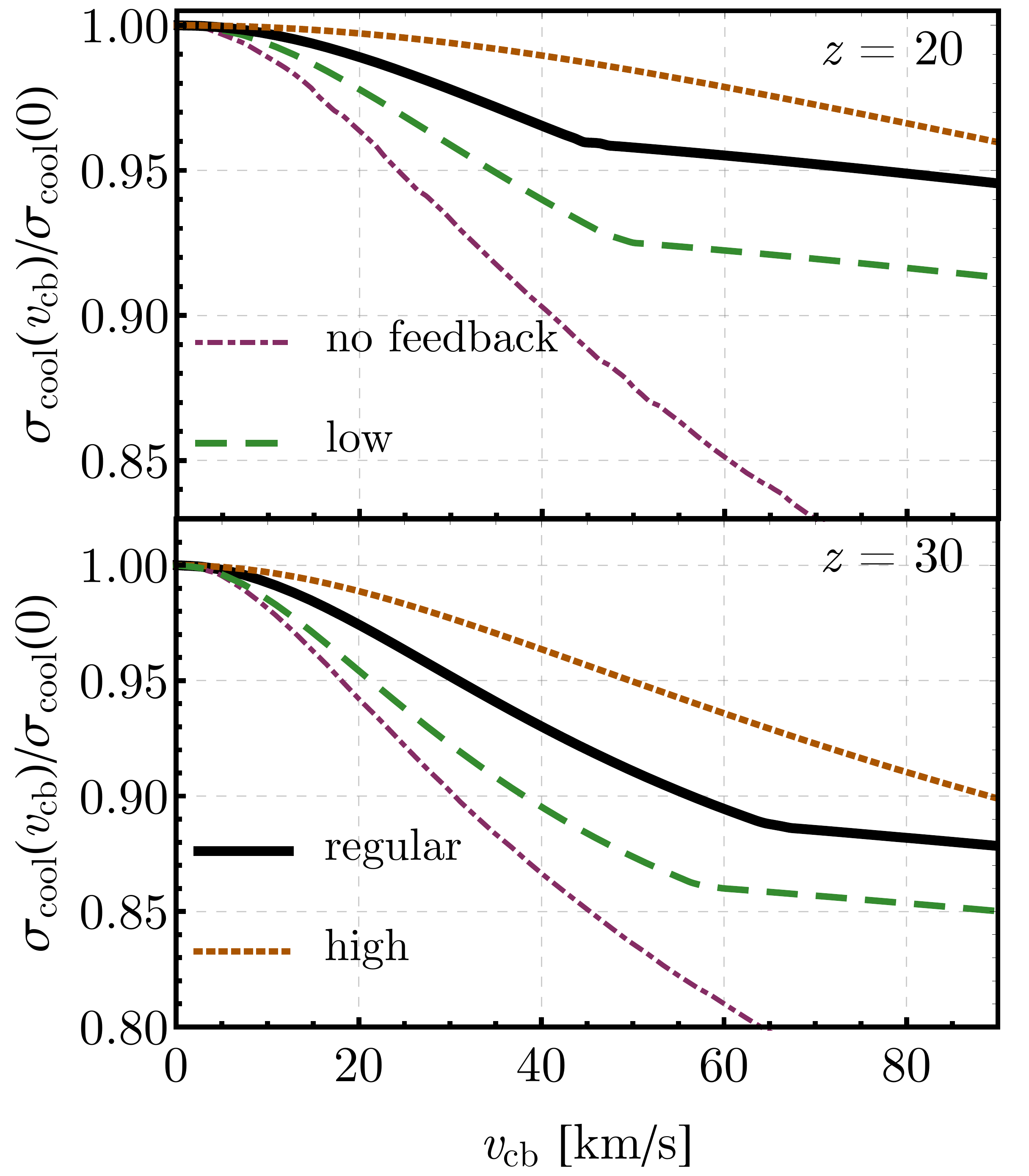}
	\caption{Standard deviation $\sigma_{\rm cool}$ of matter fluctuations in regions that will form haloes massive enough to cool and form stars, according to Eq.~\eqref{eq:McoolLW}, at $z=20$ (top panel) and $z=30$ (bottom panel), normalized.
		Each line represents a different feedback assumption, as defined in Section~\ref{sec:vcbeffect}.
		The kinks in the regular and low-feedback curves appear due to the transition from molecular- to atomic-cooling haloes, where one of the $\vcb$ effects vanishes.
	}
	\label{fig:sigmacool}
\end{figure}

We show the standard deviation of overdensities in star-forming haloes as a function of $\vcb$, defined as $\sigma_{\rm cool}(z,\vcb) = \sigma_{M}(z,\vcb)$ for $M=M_{\rm cool}(z,\vcb)$,
in Fig.~\ref{fig:sigmacool}, for redshifts $z=20$ and 30.
We normalize all lines to their $\vcb=0$ value, with values of $\sigma_{\rm cool}=\{0.52, 0.42, 0.41, 0.41\}$ and $\{ 0.37, 0.32,0.31,0.31\}$, for no feedback, low, moderate, and high feedback, at $z=20$ and 30, respectively.
This figure shows that, in all cases, higher velocities require larger haloes to form stars, translating into lower values of $\sigma_{\rm cool}$.
The size of this effect decreases with the strength of the LW feedback, as it raises the minimum cooling mass, and increases with redshift, as the velocities are more relevant at earlier times, although the size of the effect does not scale simply as $\vcb(z)$, as it depends on its integrated history.
Finally, we note that the ``kinks" in some of the lines with feedback in Fig.~\ref{fig:sigmacool} correspond to the transition to atomic-cooling haloes, where the effect of $\vcb$ in $V_{\rm cool}$ (as in Eq.~\eqref{eq:vcool}) vanishes.

\begin{figure}[htbp]
	\includegraphics[width=0.5\textwidth]{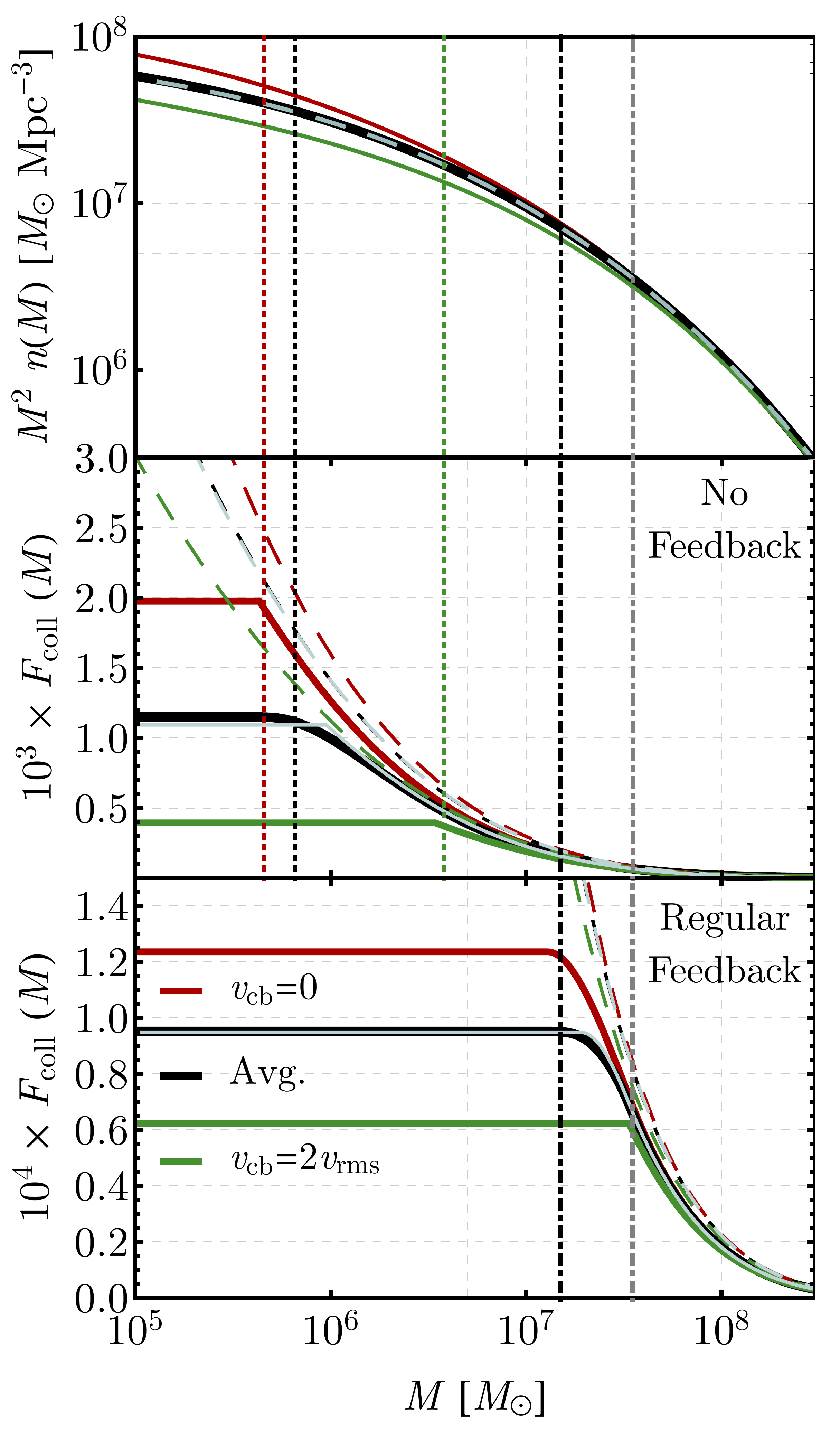}
	\caption{In the upper panel we show the HMF, and in the bottom panels the collapsed fraction of gas (in solid line) and all matter (in dashed lines) as a function of the minimum halo mass $M$ considered, for the cases of no feedback in the middle panel and regular feedback in the bottom, all at $z=20$.
	We show the results for $\vcb=0$ in red, for $2 \, v_{\rm rms}$ in green, and the average over velocities in black, where it is clear that larger relative velocities correspond to smaller collapsed fractions.
	We also overlay the $\vcb=v_{\rm rms}$ case in the thin blue line, which is almost coincident with the average case.
	Finally, the vertical colored lines correspond to the minimum mass necessary to cool and form stars at this redshift for different assumptions, where
		each colored dotted line corresponds to a $\vcb$ case mentioned above (assuming no feedback), whereas the dot-dashed lines represent the minimum cooling mass in the regular feedback case (black, for $\vcb=0$) and for atomic cooling (gray).
	}
	\label{fig:combinedHMF}
\end{figure}

When computing the HMF we work with the dimensionless quantity~\cite{Sheth:1999su,Naoz:2006tr}
\be
\nu (M,z, \vcb) = \left( \dfrac{\delta_{\rm crit}}{\sigma_M(z,\vcb)}\right)^2,
\ee
where $\delta_{\rm crit} \equiv 1.69 - \gamma\,(1+z)$ is the critical overdensity for collapse, with $\gamma=3\times 10^{-5}$ serving as a small correction to account for the effect of baryons at high redshifts~\cite{Naoz:2006ye}.
The Reed HMF~\cite{Reed:2006rw} has the following form:
\be
\nu f(\nu) = \mathcal A  (1+\beta_1 \nu'^{-p} + \beta_G\, G_1) \sqrt{\nu'} \exp(-\nu'/2.0),
\ee
with $\beta_1 = 1.02$, $\beta_G = 0.2$, $p=0.3$, 
\be
G_1 = \exp \left[{ - \dfrac{(\log(\nu')-d)^2}{2\times (0.6)^2}}\right],
\ee
$d=0.788$, and $\nu'= a \nu$ with $a=0.764$.
The amplitude $\mathcal A$ is found by normalizing the HMF at all masses, so that $\int d\nu f(\nu) = 1$.
Then, we find the HMF as a function of velocity to be
\be
n(M,z,\vcb) = \dfrac{\rho_m^0}{M^2} \left(-2\dfrac{d \log \sigma_M}{d \log M}\right) \nu f(\nu).
\ee
We show this quantity (multiplied by $M^2$ to obtain the fraction of matter in haloes per unit log-mass) in Fig.~\ref{fig:combinedHMF}, for three values of the relative velocity, where it is clear that regions with larger relative velocities form fewer small-mass haloes, whereas for atomic-cooling haloes with $M\gtrsim {\rm few}\times 10^7\,M_\odot$ the difference is less significant.

\subsection{All Combined}

Finally, we can combine all the aforementioned effects to calculate the fraction of baryons that collapse to star-forming haloes as~\cite{Visbal:2012aw}
\be
F_{\rm coll} (z,\vcb) = \int_{M_{\rm cool}}^\infty\!\!\!\!\!\!\!\! dM M n(M) \dfrac{f_{\rm gas}}{\rho_b} f_*,
\label{eq:Fcoll}
\ee
where $\rho_b$ is the comoving baryon abundance.
This includes the velocity effect on the HMF (through $n(M)$); on the gas fraction $f_{\rm gas}$ of haloes; and on the minimum cooling mass $M_{\rm cool}$, all of which are functions of redshift and $\vcb$.
To illustrate the dependencies in this function, we show $F_{\rm coll}$ as a function of the minimum cooling mass $M$ in Fig.~\ref{fig:combinedHMF},  for both the no-feedback and the regular-feedback cases.
There we see how larger velocities decrease the number of haloes, the gas fraction in every halo, and increase the minimum cooling mass. These effects are less apparent in the case with feedback, although still present to a large degree.

\begin{figure}[hbtp]
\includegraphics[width=0.5\textwidth]{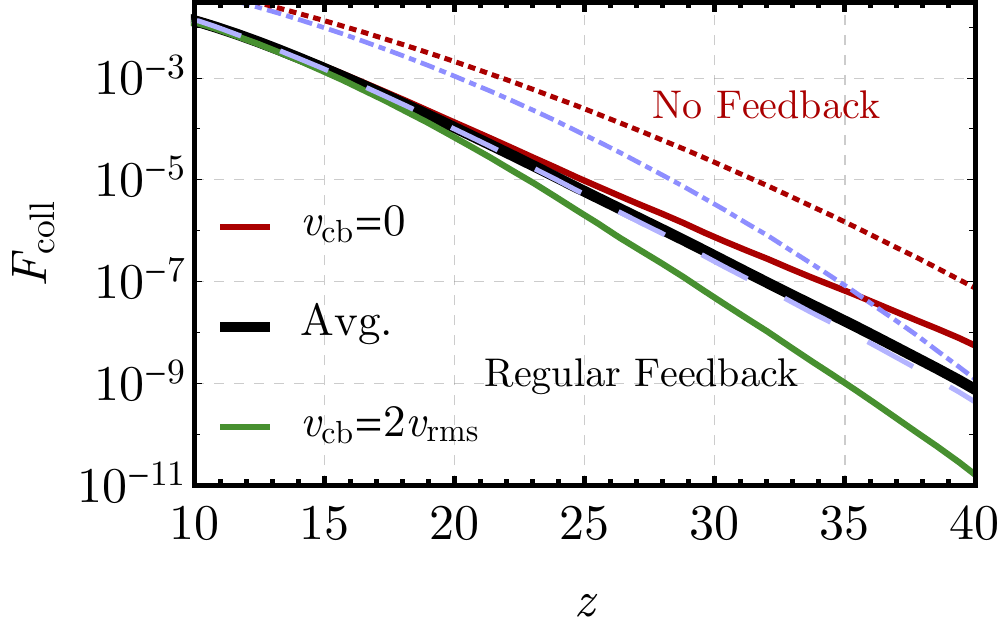}
\caption{We show the collapsed fraction of baryons to star-forming haloes as a function of redshift, for our regular feedback case, for patches with $\vcb=0$ in the solid red line, $\vcb=2\,v_{\rm rms}$ in the green line, and for the average over the sky in the black line.
Additionally, we show the $\vcb = v_{\rm rms}$ case in the light-blue long-dashed line, almost coincident with the average case.
This plot includes the three effects of $\vcb$ in the formation of stars, through the HMF, the gas fraction in haloes, and the minimum halo mass to form stars.
For comparison, the red dotted and blue dash-dotted lines show the $\vcb=0$ and $\vcb=v_{\rm rms}$ results for the no-feedback case.
In all feedback cases patches with large $\vcb$ have a smaller collapsed fraction, and thus fewer stars.
}
\label{fig:plotFcoll}
\end{figure}

We show the collapsed function of baryons that can form stars, from Eq.~\eqref{eq:Fcoll}, in Fig.~\ref{fig:plotFcoll} as a function of redshift, for different values of $\vcb$, as well as its average over velocities.
From this figure we see that velocities have a larger impact at higher $z$, as their effects partially redshift away at later times, with the average $F_{\rm coll}$ being a factor of 2-5 smaller than the $\vcb=0$ case.
This illustrates the necessity of accounting for relative velocities if one is interested in molecular-cooling haloes, since otherwise one would form too many of them. This can be solved easily with {\tt 21cmvFAST}.

As we will see, our observable---the 21-cm brightness temperature---depends on $F_{\rm coll}$ nonlocally, as photons emitted from a galaxy can travel far from its source.
The approach of {\tt 21cmFAST} to solve this problem is to use an excursion set approach~\cite{Furlanetto:2004nh,Mesinger:2007pd} to compute the amount of photons emitted at earlier times from distant sources. 
For that the code smooths the density field in spherical regions of different radii $R$ around each cell, and uses the averaged field to compute the collapse fraction, which tells us how many photons were emitted from this region.
Computationally we use the extended Press-Schechter (EPS) formalism to account for over- and under-densities on top of velocity fluctuations~\cite{Barkana:2003qk}.
Assuming that we take a patch of radius $R$, with averaged overdensity  $\delta$ and relative velocity $\vcb$, and $\sigma_R^2$ is the matter density variance in patches of radius $R$, we follow Refs.~\cite{Barkana:2003qk,Tseliakhovich:2010yw} in writing the collapsed fraction of baryons in this patch as
\be
F^{\rm tot}_{\rm coll}(\delta,\vcb) = \dfrac{F_{\rm coll}(\vcb)}{\overline{F_{\rm EPS}}} F_{\rm EPS}(\delta,\vcb),
\ee
where $F_{\rm EPS}(\delta,\vcb)$ is the EPS collapsed fraction, given by
\be
F_{\rm EPS}(\delta, \vcb) = {\rm erfc} \left [ \dfrac{\delta_{\rm crit} -  \delta}{\sqrt{2[\sigma_{\rm cool}^2( \vcb) - \sigma_R^2]}}\right],
\label{eq:EPS}
\ee
where erfc is the complementary error function, and $\overline{F_{\rm EPS}}$ is the global average of $F_{\rm EPS}$.
We follow the prescription in {\tt 21cmFAST} by taking the nonlinear overdensities in $\delta$ (i.e., extrapolating the $z=0$ result backwards), and calculating $\overline{F_{\rm EPS}}$ numerically by averaging over our cells~\cite{Mesinger:2010ne,Greig:2015qca}.

Notice that in the limit $R\to\infty$, the collapsed fraction in Eq.~\eqref{eq:EPS} asymptotes to $F_{\rm EPS}(0,\VEV{\vcb})$, which is not necessarily the same as $\VEV{F_{\rm EPS}(\delta,\vcb)}$ (where $\VEV{ }$ denotes spatial average). 
Nonetheless, as we divide by the actual averaged value of $\overline{F_{\rm EPS}}$, any bias that could arise from this procedure is cancelled out.
Additionally, as shown in Figs.~\ref{fig:combinedHMF} and~\ref{fig:plotFcoll}, the averaged $F_{\rm coll}$ is almost identical to its value for $\vcb=\VEV{\vcb}$.
This is because the probability density function (PDF) for $v_{\rm cb}^2$ is highly peaked around $v_{\rm rms}^2$, where $v_{\rm rms}\approx 30$ km s$^{-1}$ is the root mean square of the Maxwell-Boltzmann distribution of $\vcb$, so using the mode value gives very similar results to averaging over the PDF.
Additionally, for computational efficiency we find the $\sigma_{\rm cool}(\vcb)$ (as plotted in Fig.~\ref{fig:sigmacool}) by calculating $F_{\rm coll}$ with a Press-Schechter HMF~\cite{Press:1973iz} including all velocity effects,
as it takes the simple form in Eq.~\eqref{eq:EPS} for $\delta=\sigma_R=0$, and one can find $\sigma_{\rm cool}(\vcb)$ by inverting the erfc function.

\begin{figure*}[hbtp]
	\includegraphics[width=0.7\textwidth]{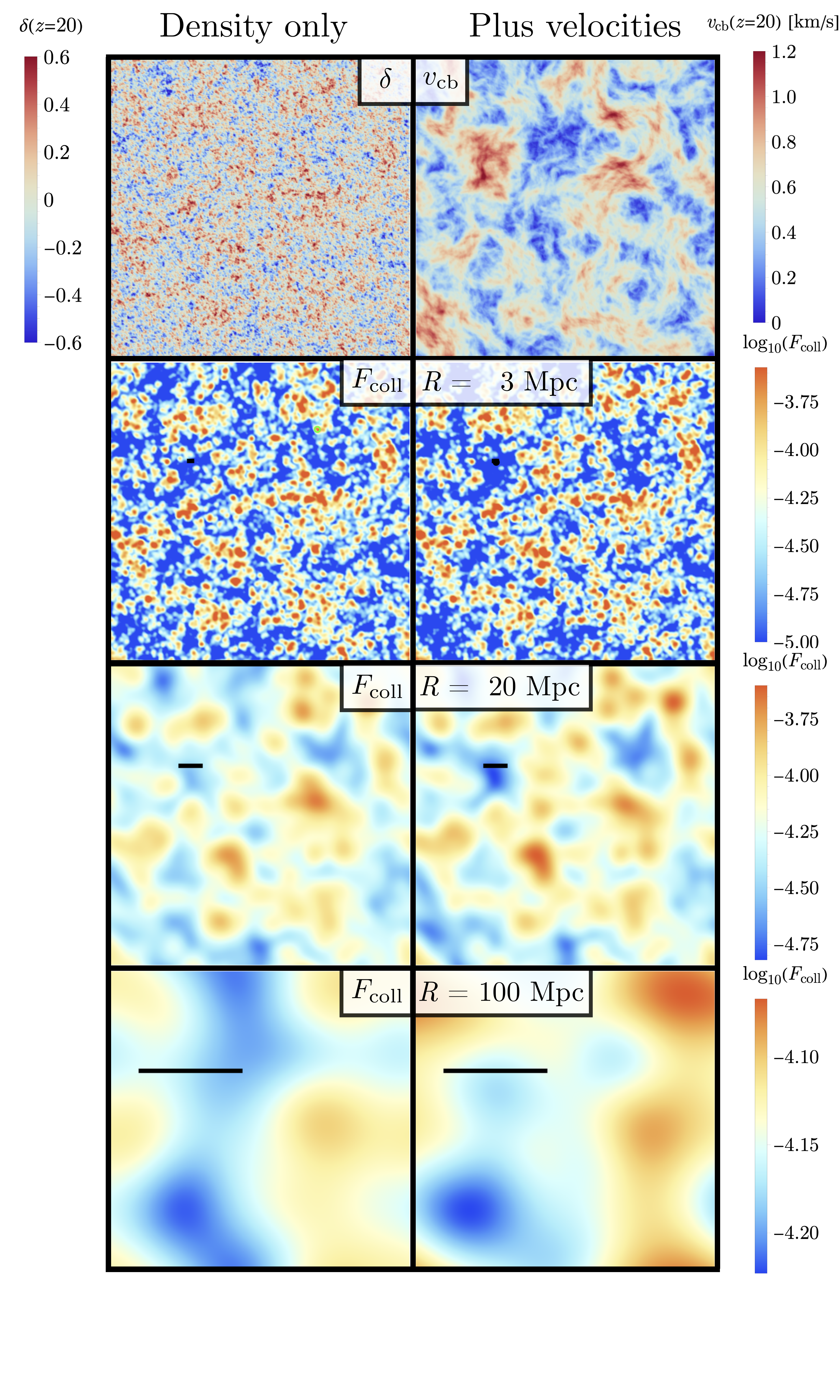}
 \caption{Collapsed fraction ($F_{\rm coll}$) without (\emph{left}) and with (\emph{right}) the effect of DM-baryon relative velocities ($\vcb$) at $z=20$, smoothed over different radii $R$, which shows how the fluctuations on $\vcb$ are imprinted onto $F_{\rm coll}$.
  The top panels show a slice 1 Mpc thick with a 300 Mpc side of a simulation box, generated with {\tt 21cmvFAST}, where in the left we show the over/underdensities in matter, and on the right the fluctuations in the DM-baryon relative velocity $\vcb$, both quantities evaluated at $z=20$.
  The second and third rows show the fraction $F_{\rm coll}$ of baryons collapsed to star-forming galaxies, as defined by Eq.~\eqref{eq:Fcoll}, smoothed over $R= 3$ and $20$ Mpc, respectively, as typical of the mean-free path of X-ray photons with 0.2 and 0.5 keV energies~\cite{Furlanetto:2009uf}.
 The final row is also $F_{\rm coll}$, smoothed over distances of 100 Mpc, corresponding to the typical travel distance of a UV photon until it enters the Lyman-$\alpha$ resonance~\cite{Dalal:2010yt}. 
 Note that in the left panels we have fixed $v_{\rm cb} = \overline{v_{\rm cb} }$, as opposed to $v_{\rm cb} =0$, in order to keep roughly the same average collapse fraction, the color scales are the same in both left and right $F_{\rm coll}$ plots, and 
 in all cases we assume regular-strength Lyman-Werner feedback.
 For reference the horizontal black lines denote the smoothing distance $R$ in each panel.
 }
	\label{fig:jointFcoll}
\end{figure*}

\section{The 21-cm line}
\label{sec:21cmline}

The distribution of the first stars during cosmic dawn cannot be directly probed.
Fortunately, their presence can be indirectly inferred with measurements of the 21-cm line, which we now describe.

\subsection{The Basics}

We start by briefly reviewing the physics of the 21-cm hydrogen line (for a detailed review we encourage the reader to see Refs.~\cite{Furlanetto:2006jb,Pritchard:2011xb,Barkana:2016nyr}).
Hydrogen, in addition to being the most abundant element in our Universe, possesses a rich hyperfine structure. Of particular interest is its spin-flip transition from the triplet to the singlet ground state, which produces a photon with a 21-cm wavelength.
To determine whether cosmological hydrogen emits or absorbs 21-cm photons during a certain era, one has to calculate its spin temperature and compare with that of the CMB.
Given a comoving number density of hydrogen atoms in the triplet ($n_1$) and singlet ($n_0$) states, we can define the local spin temperature $T_s$ through
\be
\dfrac{n_1}{n_0} = \dfrac{g_1}{g_0} \, e^{-T_*/T_s},
\ee
where $g_1=3$ and $g_0=1$ are the numbers of degrees of freedom of the triplet and singlet states, respectively, and $T_*\approx 0.068$ K is the temperature corresponding to the 21-cm hyperfine transition.
If $T_s>T_\gamma$, the hydrogen will emit 21-cm photons, and if $T_s<T_\gamma$, it will absorb photons from the Raleigh-Jeans tail of the CMB.
We can define the brightness temperature of the 21-cm line as~\cite{Pritchard:2011xb}
\be
T_{21} = 38\,{\rm mK }\, \left(1 - \dfrac{T_\gamma}{T_s}\right) \left(\dfrac{1+z}{20}\right)^{1/2}\!\!\! x_{\rm HI} (1 + \delta_b) \dfrac{\partial_r v_r}{H(z)},
\label{eq:T21}
\ee
for our fiducial cosmological parameters, where $x_{\rm HI}$ is the neutral-hydrogen fraction (nearly unity for all eras of interest in this work), and $\partial_r v_r$ is the line-of-sight gradient of the velocity.

The standard history of the 21-cm line of hydrogen is as follows.
During the dark ages collisions coupled the spin temperature to that of the gas, producing 21-cm absorption for $z\gtrsim 30$~\cite{Loeb:2003ya}.
This process, however, became inefficient as the Universe expanded and the gas cooled further, resulting in nearly no absorption until the cosmic-dawn era.
In this epoch the first stars were formed, emitting abundant UV photons, which coupled $T_s$ to $T_{g}$ through the Wouthuysen-Field effect~\cite{Wout,Field,Hirata:2005mz}, by which Lyman-$\alpha$ photons resonantly scatter between hydrogen atoms, imprinting the kinetic temperature onto the hyperfine populations of hydrogen. 
This produces---once again, 21-cm absorption---and we call this period the Lyman-$\alpha$ coupling era (LCE).
Later on, X-ray photons will be produced in a large-enough quantity to reheat the intergalactic medium to temperatures $T_g\gg T_\gamma$, switching the 21-cm signal to emission. 
We call this period the epoch of heating (EoH).
Finally, during the epoch of reionization, at $z \lesssim 10$, there will be enough UV photons that all neutral hydrogen will be ionized, until no 21-cm signal is left.

We are interested in the 21-cm line during cosmic dawn ($z\sim 20$), when the spin temperature is set by the competition between couplings to the CMB and Lyman-$\alpha$ backgrounds, and the gas temperature is first dominated by adiabatic cooling of the gas, and then rises due to X-ray heating. 
Given the complexity of this signal, the most straightforward approach is to perform quasi-numerical simulations. 
We do so with {\tt 21cmvFAST}.

\subsection{Implementation on {\tt 21cmvFAST}}

We include the effects of relative velocities in the commonly used {\tt 21cmFAST} code~\cite{Mesinger:2010ne}, by modifying the streamlined version called {\tt 21CMMC} presented in Refs.~\cite{Greig:2015qca,Greig:2017jdj}.
Here we outline the main changes to the code, and the results found, and we encourage the reader to visit the webpage with the modified code, where all changes are detailed.

Each realization of {\tt 21cmvFAST} is composed of a box of side $L$ and cell size $R_{\rm cell}$, with a set of initial densities $\delta^{(i)}(\mathbf x)$ given by the matter power spectrum, evolved using linear or second-order Lagrangian perturbation theory to lower redshifts~\cite{Scoccimarro:1997gr}.
This is an excellent approximation for the high redshifts ($z\gtrsim 10$) that we are interested in.
We will always choose $R_{\rm cell} \leq 3$ Mpc, so that our moving-background perturbation theory in Eq.~\eqref{eq:ODEs} holds.
The first modification that we implement is to generate a box of initial DM-baryon relative velocities. 
We do so (keeping correlations with over/under-densities) by assigning each Fourier-space pixel with density $\delta^{(i)} (\mathbf k)$ a velocity
\be
\mathbf \vcb^{(i)}(\mathbf k) = \dfrac{i \mathbf k}{k} \sqrt{\dfrac{P_v(k)}{P_\delta(k,z_i)}} \delta^{(i)} (\mathbf k),
\ee
where $P_v$ and $P_\delta$ are the power spectra for $\vcb$ and $\delta$, respectively,
which we Fourier transform to real space, to find $\mathbf \vcb(\mathbf x)$ at each pixel (and thus $\vcb(\mathbf x)$ by taking its norm). 
We show a slice of the initial conditions in our simulation in Fig.~\ref{fig:jointFcoll}, where it is clear that overdensities fluctuate on all scales, whereas the relative velocities are clearly coherent for scales below $\sim 10$ Mpc, and most of their power lies in BAO scales ($R \approx 150$ Mpc).

We use our calculations from Section~\ref{sec:vcbeffect} to find $F_{\rm coll}$ for each point in the box at $z=20$, assuming regular feedback strength and given our initial conditions, both including and neglecting the effects of the relative velocities, which we show in Fig.~\ref{fig:jointFcoll}.
This figure clearly showcases how regions with high velocity see a significant suppression in the amount of baryons collapsed to galaxies, and thus of stars formed.
We have smoothed $F_{\rm coll}$ over different radii, to represent the characteristic distance traveled by different kinds of photons.
X-ray photons can travel small distances before being absorbed, if their energies are low, so we have averaged over radii  $R=3$ and 20 Mpc, corresponding to energies of 0.2 and 0.5 keV, respectively.
On the other hand, during cosmic dawn the UV photons that produce Lyman-$\alpha$ coupling will typically travel as far as $R=100$ Mpc before redshifting into the Lyman-$\alpha$ transition, as they typically do not get absorbed.
We see from Fig.~\ref{fig:jointFcoll} how smoothing over larger $R$ partially dampens the effect of velocities at smaller scales, as expected, although even for $R=100$ Mpc there is a visible signature of $\vcb$ in the collapsed fraction, which will be inherited by the 21-cm power spectrum.

\begin{figure}[hbtp]
	\includegraphics[width=0.52\textwidth]{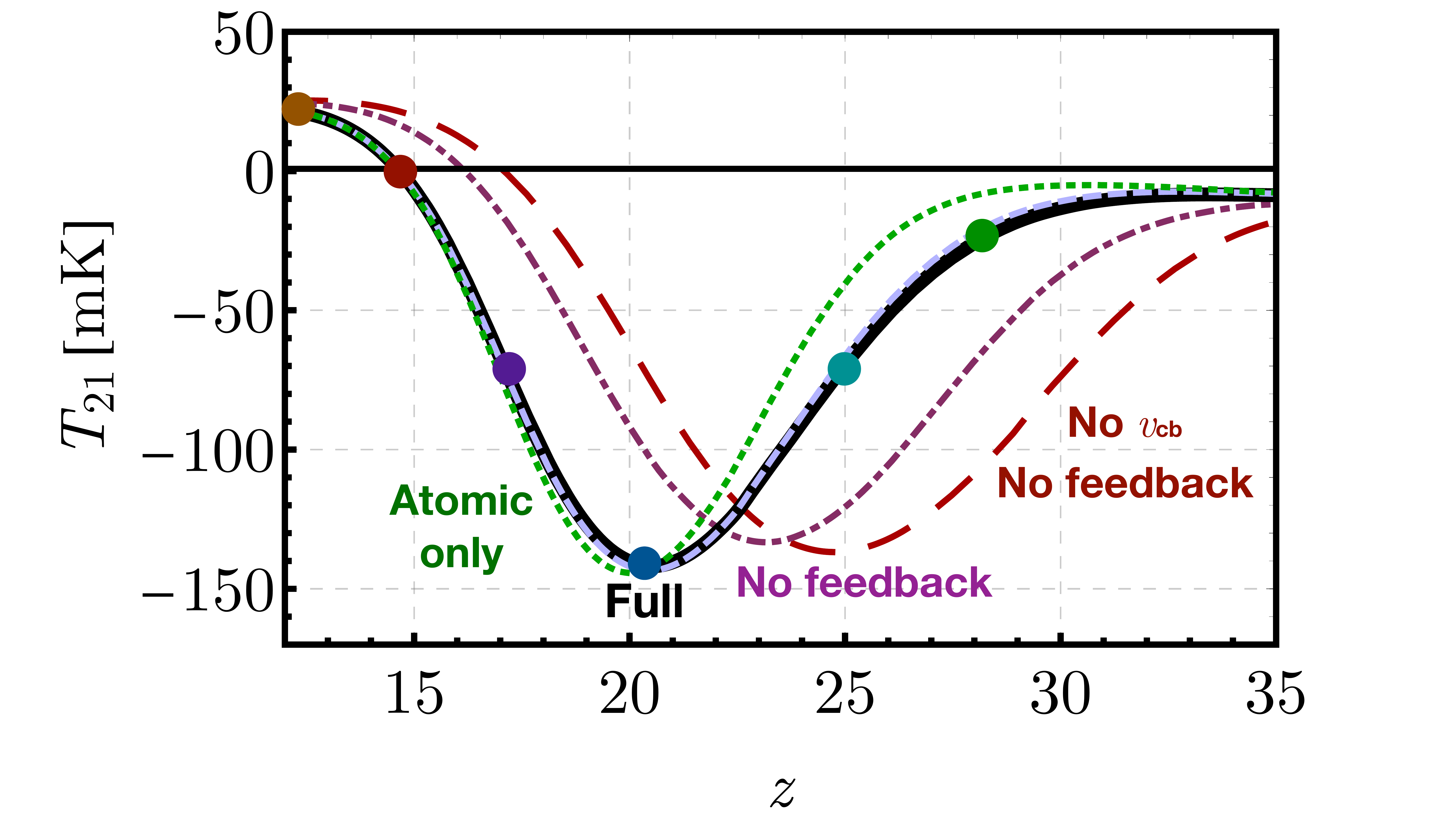}
	\caption{Sky-averaged 21-cm brightness temperature, obtained from our simulations, as a function of redshift for different assumptions.
	In the red long-dashed line we show the usual molecular-cooling case, without including the effect of velocities or Lyman-Werner feedback.
	In the purple dot-dashed line we show the results when including $\vcb$, but not feedback, and the  black heavy line shows the result of the full calculation, including both $\vcb$ and LW feedback.
	The blue dashed line represents the result when fixing $\vcb=v_{\rm avg}$, without fluctuations, whereas the green dotted line is the result if only atomic-cooling haloes form stars, as commonly assumed in the literature.
	The inclusion of both relative velocities and Lyman-Werner feedback delays the global-signal landmarks. 
	Finally, the colored circles represent the redshifts at which we will show maps of $T_{21}$ in Figs.~\ref{fig:21cmLyA} and \ref{fig:21cmXray}.
	}
	\label{fig:T21avg}
\end{figure}

\begin{figure*}[t]
	\includegraphics[width=0.99\textwidth]{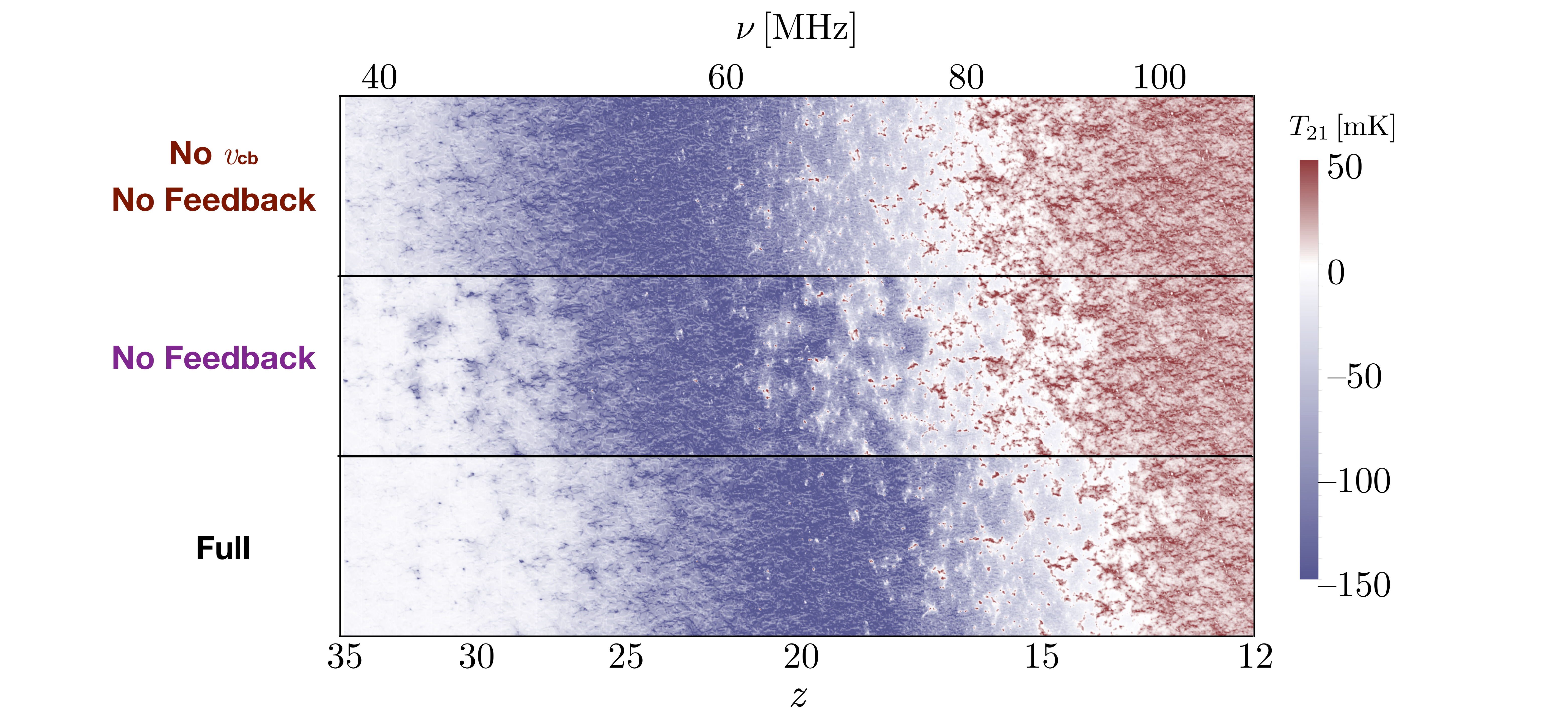}
	\caption{The 21-cm brightness temperature across cosmic dawn in the case without LW feedback or velocities (\emph{top}), with $\vcb$ but no feedback (\emph{middle}), and the full case with (regular-strength) feedback and velocities (\emph{bottom}).
	Each result is obtained from a simulation slice 300 Mpc in length and 1 Mpc in thickness, consistently evolved in redshift $z$.
	Blue and red represent 21-cm absorption and emission, respectively, and the anisotropies are primarily sourced by the formation of the first stars.
	}
	\label{fig:horizontal}
\end{figure*}
\subsection{Global Signal}

Given the implementation of {\tt 21cmvFAST} outlined above, we can now calculate the 21-cm brightness temperature for any combination of cosmological and astrophysical parameters, including molecular-cooling haloes with a time-dependent LW feedback, and the effect of relative velocities.

In all of our simulations we take an X-ray luminosity of $\log_{10}(L_X/\rm SFR)=40$ (divided by the star-formation rate SFR), and a threshold in X-ray energy of $E_0=0.2$ keV~\cite{Visbal:2012aw}, with a spectral index of $\alpha_X=1$, and do not modify the rest of the standard {\tt 21CMMC} parameters.
We will also explore the case of $E_0=0.5$ keV in App.~\ref{sec:AppXray}, as the first X-ray sources might have harder spectra than we assume~\cite{Fialkov:2014kta,Pacucci:2014wwa,Cohen:2017xpx}.
Instead of a fixed cooling temperature $T_{\rm cool,Xray}=10^4$ K for X-ray haloes we use the prescription for $F_{\rm coll}$ and $\sigma_{\rm cool}$ outlined in Section~\ref{sec:vcbeffect}.
Additionally, we have not modified the ionization part of {\tt 21cmFAST}, assuming that the haloes we study here provide a negligible amount of UV photons that escape into the IGM. 
Therefore, for ionizations we employ the ``Faint Galaxies" model of Ref.~\cite{Mesinger:2016ddl}, where reionization is driven by galaxies with $T_{\rm cool,UV}=2\times10^4$ K, with an ionizing efficiency of $\zeta=20$, and by the end of our simulations, at $z=12$, we always have $x_e \ll 1$.

We show in Fig.~\ref{fig:T21avg} the average 21-cm brightness temperature $\VEV{T_{21}}$ in our simulation boxes as a function of redshift.
The inclusion of relative velocities delays the formation of the first stars, shifting all landmarks from the ``plain" molecular-cooling case (i.e., ignoring the effects of both $\vcb$ and Lyman-Werner feedback)  $\Delta z=2$ later, in agreement with the results in Ref.~\cite{Fialkov:2012su}.
Accounting for LW feedback as well produces a further delay of $\Delta z=3$, due to the need for more-massive haloes to form stars at lower redshifts.
As a consistency check,  we also show in that plot the result for a simulation with fixed $\vcb=v_{\rm avg}=\VEV{\vcb} = 0.92 v_{\rm rms}$, where no velocity-induced fluctuations are present.
We see that in this case the average 21-cm temperature is very similar to the full case.
So while $\vcb$ affects the background evolution of the 21-cm line in a particular way, these effects can potentially be mimicked by time-dependent feedback on the star-formation rate, and are thus not unique.
We also show, for comparison, the background evolution of $T_{21}$ in the unphysical case that LW feedback is saturated at all times, and only atomic-cooling galaxies are allowed to form stars. 
We elaborate further on this case in App.~\ref{App:atomic}.

\subsection{Fluctuations and Power Spectrum}

Fortunately, there is more information in the 21-cm line than the global signal carries.
Including $\vcb$ and LW feedback delays the formation of structure, which affects the 21-cm fluctuations during the cosmic-dawn era in two ways.
First, larger velocities result in fewer Lyman-$\alpha$ photons, producing anisotropies during the LCE~\cite{Dalal:2010yt,OLeary:2012gem,Fialkov:2013jxh}.
Secondly, even after Lyman-$\alpha$ coupling is saturated, larger velocities result in fewer X-ray photons being emitted, and therefore less IGM heating~\cite{Visbal:2012aw,OLeary:2012gem}.
Following these two effects we divide the cosmic dawn into two different eras, the EoH and the LCE, which have qualitatively different fluctuation behaviors.
This is illustrated in Fig.~\ref{fig:horizontal}, where we show the 21-cm temperature in slices 1 Mpc thick of three simulation boxes (each 300 Mpc in length), across the entire cosmic dawn, starting at $z=35$ and finishing at $z=12$, for the same three cases as Fig.~\ref{fig:T21avg} (i.e., ignoring both $\vcb$ and LW feedback, ignoring only the feedback, and including the full set of effects).
Here it is very clear how the inclusion of $\vcb$ and LW feedback delays the onset of the LCE, as well as the transition to the EoH, as the first galaxies take longer to emit the necessary amount of UV and X-ray photons.
We now study the 21-cm fluctuations during each of these two eras in detail, focusing on our regular-feedback case, with the full set of feedback effects, to explore the effect of relative velocities.

Throughout this section we will employ two types of simulations.
We will show 21-cm maps with a box size of $L = 300 $ Mpc and a 1-Mpc resolution, as those are easiest to visually inspect.\footnote{We note that whenever we use small simulation boxes, with $L = 300 $ Mpc, we do so for illustration purposes only, as for small box sizes part of the $\vcb$ power at large scales is lost. For instance, for $L = 300 $ Mpc almost 10\% of the $\vcb$ power is not properly included.}
These boxes share the density and velocity fields from Fig.~\ref{fig:jointFcoll}.
When computing power spectra, however, we will increase the box size to $L = 900 $ Mpc, with a 3-Mpc resolution, in order to compute the large-scale 21-cm fluctuations with smaller variance.
In all cases  where we compare simulations with and without $\vcb$ we keep an average value $v_{\rm avg}$ for the latter, so that both cases have very similar global signals.

\subsection*{During the Lyman-$\alpha$ Coupling Era}

We start with the LCE, which stretches from the formation of the first abundant sources of Lyman-$\alpha$ photons, to the transition to X-ray heating, where $\VEV{T_{21}}$ is minimum.
This corresponds to the redshift range $z=20-28$ in our fiducial model.
During this era, higher velocities produce a shallower 21-cm signal by reducing the amount stars emitting UV photons able to produce Lyman-$\alpha$ coupling.
This is, however, a fairly nonlocal effect, as these photons can travel distances as large as 100 Mpc before transitioning into the Lyman-$\alpha$ line, which will smear the VAOs~\cite{Dalal:2010yt}.

\begin{figure}[hbtp]
	\includegraphics[width=0.5\textwidth]{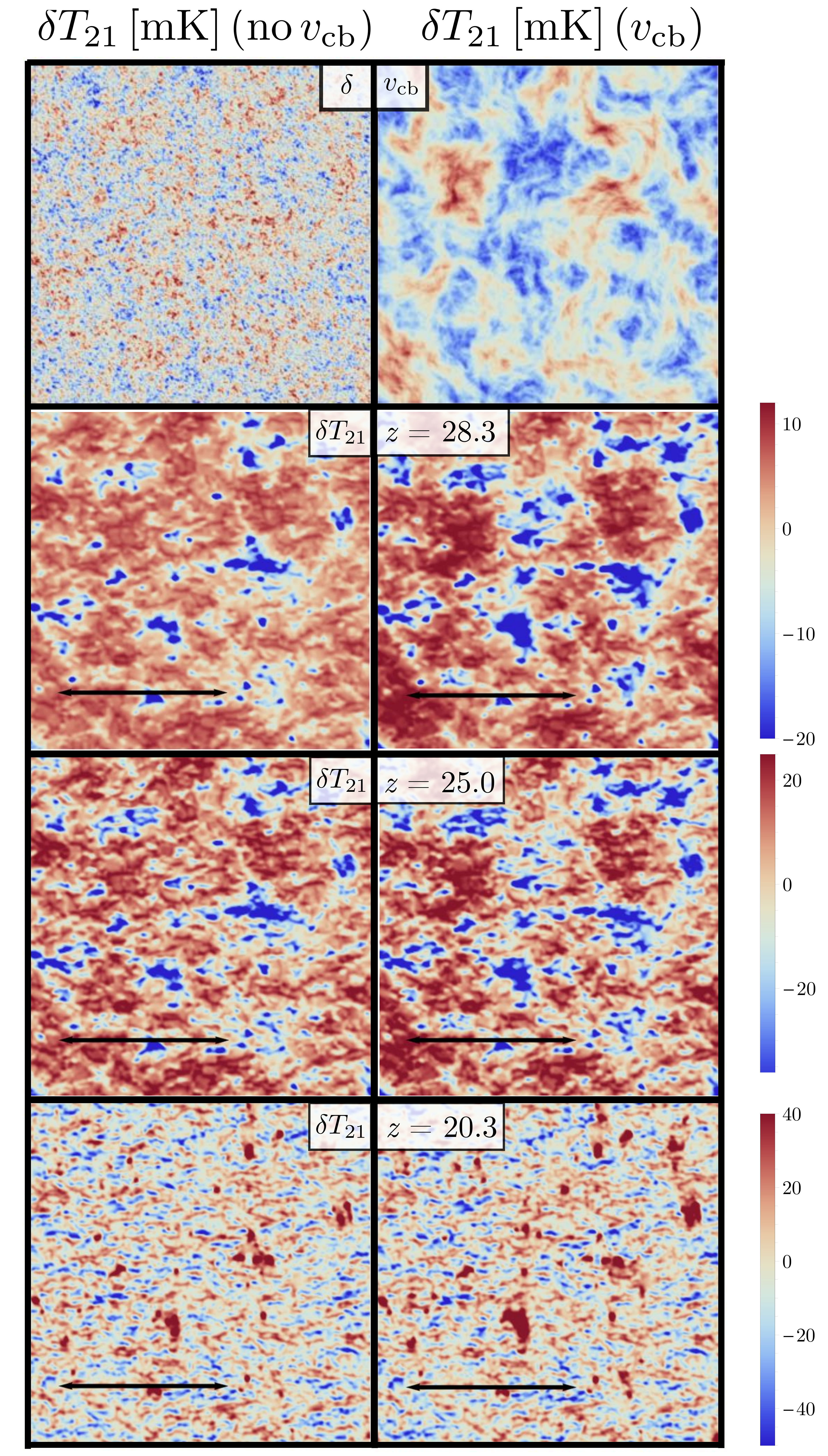}
	\caption{Fluctuations in the 21-cm brightness temperature (around their average) during the Lyman-$\alpha$ coupling era for the case of regular feedback. On the left panels we have fixed the DM-baryon relative velocity to its average value $v_{\rm avg}$, whereas on the right panels it fluctuates, with the initial conditions shown in the first row.
	The three redshifts are chosen to be on the onset of Lyman-$\alpha$ coupling ($z=28.3$), halfway through the Lyman-coupling era ($z=25.0$), and at the transition between Lyman-coupling and X-ray heating (where $\VEV{T_{21}}$ is minimum; at $z=20.3$).
	The black arrow represents the acoustic scale of 150 Mpc, for reference.
	}
	\label{fig:21cmLyA}
\end{figure}

We show 21-cm maps during the LCE in Fig.~\ref{fig:21cmLyA}, both with and without velocity fluctuations.
This Figure shows three representative redshifts, as marked by blue and green circles in Fig.~\ref{fig:T21avg}, as landmarks of the LCE.
The first redshift we show is $z=28.3$, at the onset of Lyman-$\alpha$ coupling.
Here regions of large velocity produce shallower 21-cm absorption (shown in red), whereas regions with $\vcb\approx 0$ produce deeper absorption (shown in blue).
This effect is perhaps more apparent at $z=25.0$, halfway through the Lyman-coupling era (where $\VEV{T_{21}} = 0.5\times \VEV{T_{21}}_{\rm min}$), where regions of large $\vcb$ translate into red ``islands"  in Fig.~\ref{fig:21cmLyA}.
Finally, we show the 21-cm fluctuations at $z=20.3$, in the transition between Lyman-coupling and X-ray heating (where $\VEV{T_{21}}$ is at its minimum).
Here we find negligible large-scale VAOs.
This is to be expected, since the effect of the velocities on the 21-cm brightness temperature during the EoH and LCE are opposite, and ought to cancel out at some intermediate redshift, as we will quantify further below.

The simulation boxes that we have shown so far clearly indicate that DM-baryon relative velocities are important for the evolution of the 21-cm signal. 
Nonetheless, to more precisely quantify their effect we will now study the power spectrum of the 21-cm signal.
Given the fluctuations $\delta T_{21}$ in the 21-cm temperature, we define the 21-cm power spectrum through the two-point function as usual,
\be
\VEV{\delta T_{21}(\mathbf k)\delta T^*_{21}(\mathbf k')} = (2\pi)^3 \delta_D(\mathbf k - \mathbf k') P_{21}(\mathbf k),
\ee
where $\delta_D$ is the Dirac delta function.
For simplicity we will use the amplitude of 21-cm fluctuations, defined as
\be
\Delta^2(k,z) = P_{21}(k,z) \dfrac{k^3}{2\pi^2},
\ee
where we drop the subscript ``21" and the vector symbol for $k$, as we ignore line-of-sight (LoS) anisotropies in this work.
Hereafter we refer to $\Delta^2(k)$ as the 21-cm power spectrum for simplicity.
In all cases the power spectrum of our simulations, and its errorbars, are obtained from a {\tt FFTW} of our simulation boxes, as in the standard {\tt 21cmFAST}~\cite{Frigo:2005zln,Mesinger:2010ne}, albeit with a finer $k$-binning to allow us to better distinguish the large-scale acoustic oscillations.

\begin{figure}[hbtp!]
	\includegraphics[width=0.48\textwidth]{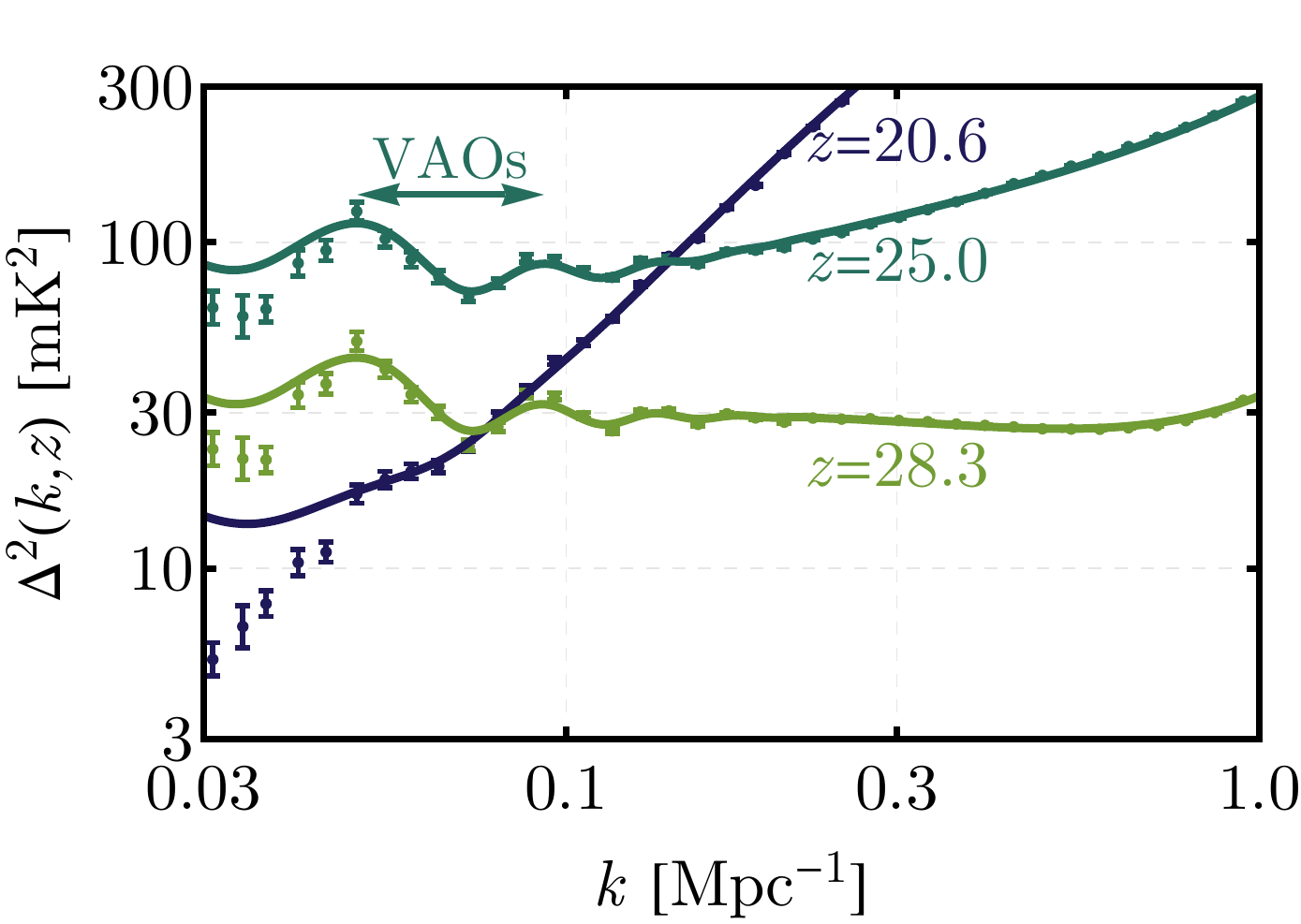}
	\caption{		
		The 21-cm power spectrum as a function of wavenumber $k$ for the same three redshifts during the LCE as Fig.~\ref{fig:21cmLyA}.
		The data points are the results of our simulations, including Poissonian error bars, whereas the solid lines are obtained with the fit in Eq.~\eqref{eq:fitpowsp} for each redshift.
		The power here is driven by the anisotropic Lyman-$\alpha$ pumping, and grows as more UV photons are emitted and couple the spin and gas temperatures.
		The arrow represents the periodicity of the VAOs, $\Delta k=2\pi/r_{\rm drag}$, given by the acoustic scale $r_{\rm drag}=147$ Mpc.
	}
	\label{fig:Pow21LyA}
\end{figure}

We show in Fig.~\ref{fig:Pow21LyA} the 21-cm power spectrum at the same three redshifts as in Fig.~\ref{fig:21cmLyA}, where we see how the fluctuations behave as a smooth function of wavenumber $k$, but at large scales are imprinted with acoustic oscillations.
These do not behave as the regular (density-induced) BAOs and are, in fact, velocity-induced acoustic oscillations (VAOs), sourced from the fluctuations on $\vcb$.
These VAOs are large at $z=25.0$, even dominating the power spectrum at large scales, but become largely irrelevant at lower $z$ as the Universe becomes fully Lyman-$\alpha$ coupled.
The size of the VAOs also decreases at higher redshifts, although as long as there is some Lyman-$\alpha$ coupling the impact of the velocities will be important.
The broadband (non-VAO) 21-cm power spectrum follows a similar evolution. 
We see in Figs.~\ref{fig:21cmLyA} and~\ref{fig:Pow21LyA} how in the transition from the LCE to the EoH all the power resides at small scales, as the first X-ray emitting sources heat up the gas around them preferentially.
We note that the regular density BAOs are included through the total-matter fluctuation $\delta^{(i)}(\mathbf k)$, whose transfer function we obtain from {\tt CLASS}~\cite{Blas:2011rf}, but their amplitude is too small (and smoothed by the complicated astrophysics of the cosmic dawn) to be observable in our power spectra.

\subsection*{During the Epoch of Heating}

We now move to the EoH, which we define to begin at the minimum of $\VEV{T_{21}}$, and to last until $T_s\gg T_\gamma$, where the entire box emits in 21 cm  ($T_{21}>0$).
For our fiducial parameters the EoH extends from $z=20$ to $z=15$, and throughout the entire EoH $T_g$ and $T_s$ are very efficiently coupled.
During this era the 21-cm fluctuations are dominated by gas-temperature anisotropies due to the inhomogeneous X-ray heating of the IGM~\cite{Pritchard:2006sq,Mesinger:2012ys}.
The way the relative velocities will impact the 21-cm signal is, then, by changing the amount of X-ray heating in different regions.
Patches with large velocities will form fewer stars, producing less heating, and therefore causing deeper 21-cm absorption.

\begin{figure}[b!]
	\includegraphics[width=0.52\textwidth]{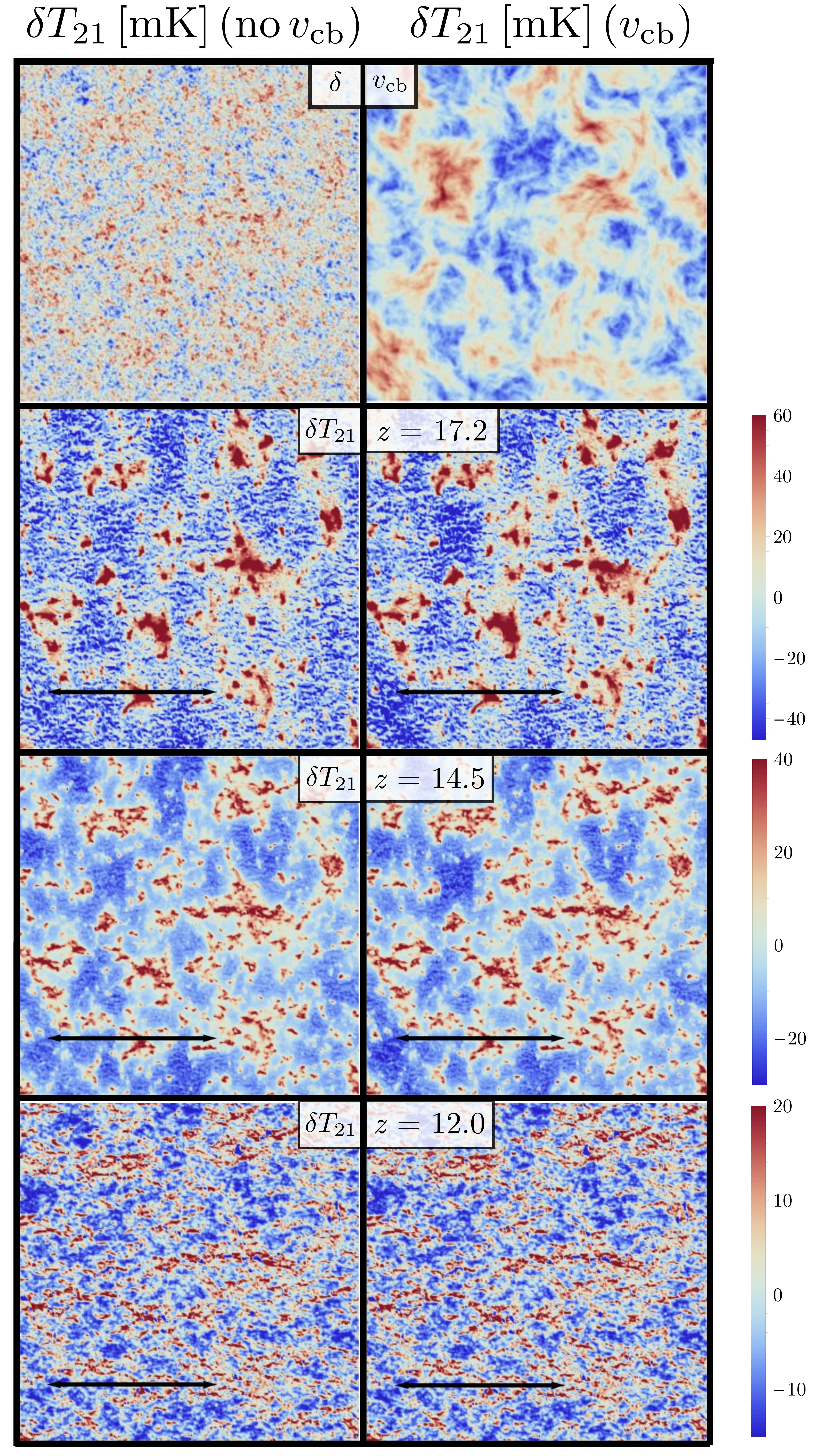}
	\caption{The 21-cm fluctuations, as in Fig.~\ref{fig:21cmLyA}, but during the epoch of heating (EoH).
	In this case we choose the redshifts to be 
    halfway through the EoH ($z=17.2$), at the transition to emission (where $\VEV{T_{21}}=0$; at $z_0=14.5$), and at the end of our simulation ($z=12.0$), where X-ray heating has saturated.
    The impact of $\vcb$ during the EoH is apparent, for instance in the third row, where larger velocities produce deeper 21-cm absorption.
	}
	\label{fig:21cmXray}
\end{figure}

As in the LCE case we will show results for three representative redshifts, now in Fig.~\ref{fig:21cmXray}.
The first is $z=17.2$, halfway through the heating transition (where $\VEV{T_{21}}=0.5\times\VEV{T_{21}}_{\rm min}$).
In this case there are marked 21-cm fluctuations at large scales due to the DM-baryon relative velocity, since regions with large relative velocities produce a more-negative $T_{21}$.
The same is true at the transition to emission, defined as $\VEV{T_{21}}=0$, which corresponds to $z_0=14.5$ in our model.
The last case we show is at $z=12.0$, at the end of our simulations, at which point the IGM has been heated to $T_s \approx T_g \gg T_\gamma$, making the $T_s$ term of Eq.~\eqref{eq:T21} negligible.
In this case clearly the DM-baryon relative velocity does not have an important effect, as $T_{21}$ has saturated.
The smooth (non-VAO) power follows roughly the same trajectory, as it is driven by the anisotropies in the distribution of stars, and thus of X-ray photons.

\begin{figure}[hbtp!]
	\includegraphics[width=0.48\textwidth]{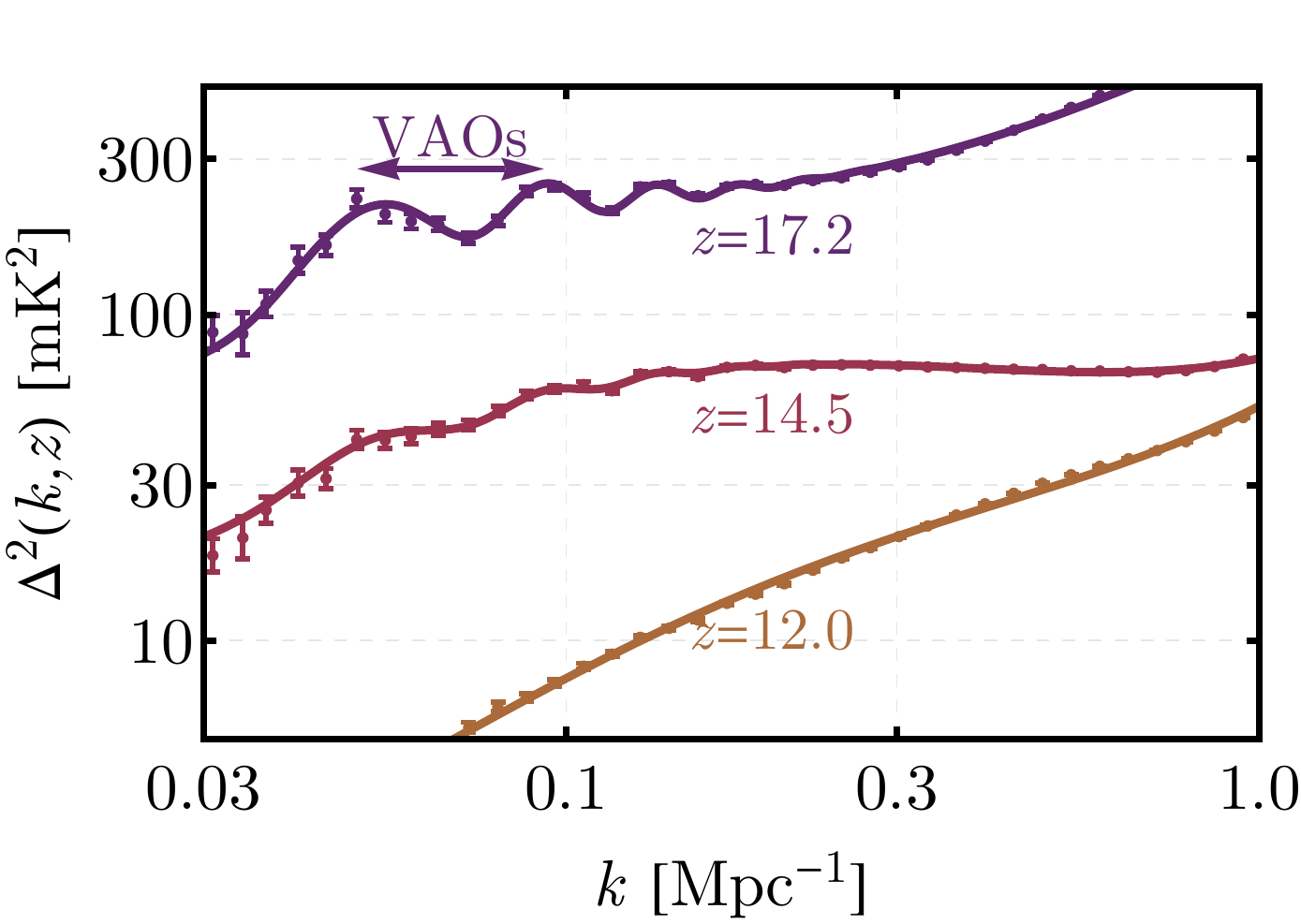}
	\caption{
		Same as Fig.~\ref{fig:Pow21LyA} but during the EoH, for the same three redshifts as Fig.~\ref{fig:21cmXray}.
		Again our simulations are shown by datapoints with error bars, and the fits are solid lines.
		The source of power here is the inhomogeneous X-ray flux of the first stars, which heats up the hydrogen gas, reducing the amount of 21-cm absorption at lower redshifts.
	}
	\label{fig:Pow21Xray}
\end{figure}

As in the previous section we plot the 21-cm power spectrum during the EoH in Fig.~\ref{fig:Pow21Xray}.
For the highest redshift, $z=17.2$, we find very important VAOs at $k\sim 0.1$ Mpc$^{-1}$, much larger than during the LCE and not severely damped.
As time progresses, X-rays slowly heat up the IGM, making the power spectrum smaller (and the VAOs less pronounced) at $z=14.5$.
We do not find a trace of VAOs in the $z=12.0$ case, as expected, since in our model $T_{21}$ has saturated by then.

Notice that here, and throughout this work, we ignore any explicit LoS anistropies in the signal. There are, however, some effects that can make the signal anisotropic, such as redshift-space distortions~\cite{Barkana:2004zy,Bharadwaj:2004nr} and the effect of the lightcone~\cite{Barkana:2005jr,Datta:2011hv} (as we only take co-evaluated boxes).
These effects can be sizable on large scales~\cite{Majumdar:2016xbu}, and potentially affect recovery of parameters from data~\cite{Greig:2018hja,Pober:2014lva}.
Nonetheless, the VAO signal is expected to be isotropic, so we leave detailed studies of the effect of LoS anisotropies for future work.

To summarize, 
at small scales ($k \gtrsim 0.3\,\rm Mpc^{-1}$) the velocities have little direct effect on the power spectrum (although of course they indirectly affect it by changing the background).
Nonetheless, on large scales ($k=0.03-0.3\,\rm Mpc^{-1}$) the 21-cm signal presents distinct VAOs, produced by the effect of the DM-baryon relative velocities on the first galaxies.

\section{Quantifying the Effect of $\vcb$}
\label{sec:quantifyvcb}

\begin{figure}[hbtp!]
	\includegraphics[width=0.48\textwidth]{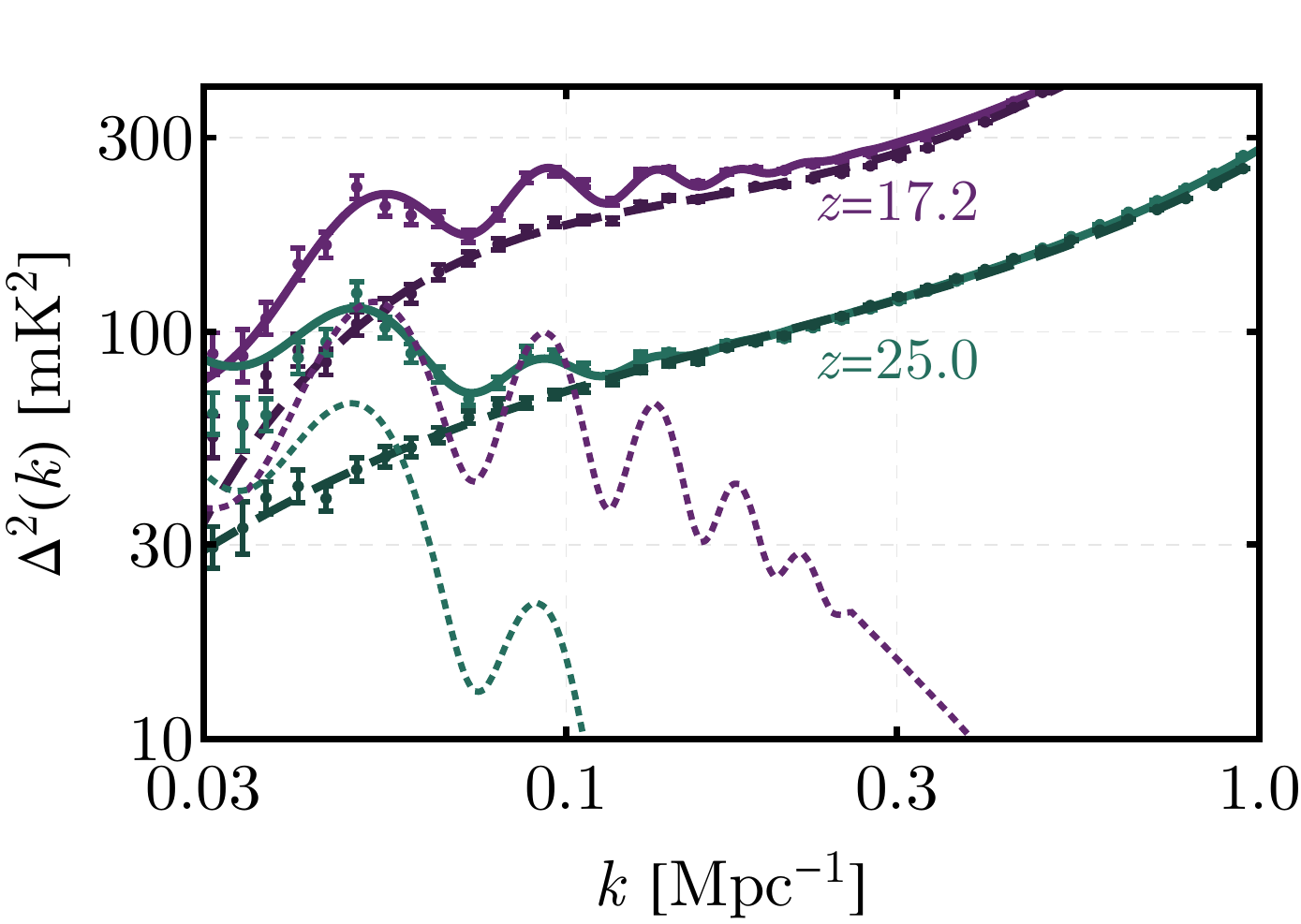}
	\caption{The 21-cm power spectrum halfway through the EoH (at $z=17.2$) and the LCE (at $z=25.0$) for our regular-strength feedback model.
	The lighter colors and solid lines represent the case with velocity fluctuations, whereas the dark colors and dashed lines show the case with DM-baryon velocities fixed at $v_{\rm avg}$, without fluctuations.
	As before, datapoints with Poissonian errors are the results of our simulations, and the lines fit them using Eq.~\eqref{eq:fitpowsp}.
	The dotted lines show the VAO-only contribution at both redshifts, which  can be added to the $v_{\rm avg}$ case to recover the total signal.
	}
	\label{fig:Pow21comparison}
\end{figure}

So far we have established that the DM-baryon relative velocities affect the 21-cm signal during the entire cosmic-dawn era, even when accounting for Lyman-Werner feedback, producing striking VAOs on large scales.
Let us, however, confirm that these acoustic oscillations are indeed produced by the relative velocities, and not over/under-densities.
For that, we perform simulations in which we fix $\vcb$ to its average value, $v_{\rm avg}$, and we compare in Fig.~\ref{fig:Pow21comparison} the 21-cm power spectrum with our regular case at two redshifts, $z=17.2$ (halfway through the EoH) and $z=25.0$ (halfway through the LCE).
This choice of a nonfluctuating $\vcb=v_{\rm avg}$ keeps roughly the same 21-cm global signal, as well as small-scale power spectrum.
Nonetheless, as is clear from Fig.~\ref{fig:Pow21comparison}, the addition of velocity fluctuations produces additional power on large scales, with acoustic wiggles.
This is what we call the VAOs.
We now move on to quantify the shape and amplitude of these VAOs, comparing them with analytic expectations.

\subsection{Shape of the Effect}

The streaming velocities fluctuate with a power spectrum sourced by the BAOs~\cite{Tseliakhovich:2010bj}, so that the power spectrum of any function $f(\vcb)$ (such as $F_{\rm coll}$ or $\delta T_{21}$) will have a component sourced by $\vcb$ fluctuations.
Two facts will greatly simplify our discussion about velocities moving forward.
First, for several examples it was shown in Refs.~\cite{Dalal:2010yt,Ali-Haimoud:2013hpa,Munoz:2018jwq} that 
\be
\VEV{f(\mathbf k)f^*(\mathbf k')} \approx (2\pi)^3 \delta_D(\mathbf k - \mathbf k') b_f^2 \Delta^2_{v^2} \dfrac{k^3}{2\pi^2},
\label{eq:powspf}
\ee
where $b_f$ is a ($k$-independent) bias, and $\Delta^2_{v^2}$ is the power spectrum of the quantity
\be
\delta_{v^2} = \sqrt{\dfrac{3}{2}} \left (\dfrac{\vcb^2}{v_{\rm rms}^2} - 1 \right),
\ee
which has no units, zero mean, and unit variance\footnote{Whenever we calculate $\vcb$ or $\Delta^2_{v^2}$ it will always be at the drag era, unless otherwise specified.}.
This approximation is, in fact, accurate to the percent level for smooth functions $f(\vcb)$, as we confirm in App.~\ref{sec:Appvsq}.
Secondly, density and relative-velocity fluctuations are uncorrelated to first order, so we can simply add the 21-cm power spectrum induced by $\delta_{v^2}$ to the one due to density fluctuations to find the total signal.
Given these two conditions, finding the impact of DM-baryon relative velocities on the 21-cm power spectrum amounts to calculating the shape and size of the term proportional to $\Delta^2_{v^2}$, which we have dubbed the VAOs.
These two conditions additionally allow us to use the VAOs as a standard ruler during cosmic dawn, as we explore in Ref.~\cite{PaperII}.

Notice that in Fig.~\ref{fig:Pow21comparison} the VAOs  do not have the same shape during the EoH ($z=17.2$) and the LCE ($z=25.0$), as in the latter case they are damped on small scales ($k \gtrsim 0.1\,\rm Mpc^{-1}$).
This is due to the propagation of Lyman-$\alpha$ photons, which typically travel much longer paths than their X-ray counterparts, erasing small-scale $\vcb$ fluctuations, as predicted in Ref.~\cite{Dalal:2010yt}.
We will detail how we include this effect for both X-ray and Lyman-$\alpha$ photons.
We will follow an approach similar to Refs.~\cite{Dalal:2010yt,McQuinn:2012rt}.

\subsubsection{Lyman-$\alpha$ Pumping}

The number $n_i$ of X-ray or Lyman-$\alpha$ photons present at a point $\mathbf x$ at redshift $z$ depends on previous redshifts $z'$ and positions $\mathbf x'$ in the past lightcone.
In particular, for Lyman-$\alpha$ photons this number is given by the integral over all previous redshifts
\be
n_\alpha (\mathbf x, z) = N_\alpha \int d^3\mathbf r d\eta \dfrac{dt}{d\eta} \dot F_{\rm coll} \dfrac{\delta_D(r - c\, \eta)}{4\pi r^2},
\label{eq:nalpha}
\ee
assuming a flat spectrum of UV photons, where $\eta$ is conformal time, $N_\alpha$ is the number of Lyman-$\alpha$ photons emitted per collapsed baryon,
and $c$ is the speed of light.
Therefore, a fluctuation in $\dot F_{\rm coll}$ due to velocities will affect the number of photons up to a significant distance away.
This nonlocality can be recast as a window function in Fourier space, where now the fluctuation in the number of Lyman-$\alpha$ photons can be written as
\be
\delta_\alpha (\mathbf k,z) = W_\alpha(k,z) \delta_F(\mathbf k,z),
\ee
where $\delta_F(\mathbf k,z)$ is the fluctuation in $F_{\rm coll}$, 
and $W_\alpha(k,z)$ is the window function, which we calculate as~\cite{Dalal:2010yt}
\be
W_\alpha(k,z) = \int_z^{z_{\rm hor}} dz' \dfrac{d F_{\rm coll}}{dz'} \sinc\left[\dfrac{k (z'-z)}{H(z')}\right],
\label{eq:windowLyA}
\ee
where $\sinc(x) = \sin(x)/x$, and we integrate up to the horizon redshift for Lyman-$\alpha$ absorptions, given by $z_{\rm hor}(z) = (32/27)(1+z)-1$, as photons from higher redshifts (with correspondingly higher frequencies) would have been absorbed as Lyman-$\beta$~\cite{Barkana:2000fd}.
Here, and throughout this discussion, we have taken the simplifying assumption that $z'\sim z$, so that $\eta-\eta'\approx (z'-z)/H(z')$,
and we show the resulting window function in Fig.~\ref{fig:windowfunc}, which  clearly damps fluctuations in Lyman-$\alpha$ photons for wavenumbers larger than $k\sim0.01$ Mpc$^{-1}$, due to the large mean-free path of these photons.
Nonetheless, the window still shows some power, albeit highly suppressed, at $k=0.1$ Mpc$^{-1}$, signaling that  Lyman-$\alpha$ pumping can potentially produce 21-cm fluctuations on scales that can be reached by upcoming 21-cm observatories.

\subsubsection{X-ray Heating}

For the case of X-rays the situation is slightly different, as their typical mean-free path is significantly shorter than that of Lyman-$\alpha$ photons. We can, nonetheless, define a window function for this scenario as well.
We will follow the calculation in {\tt 21cmFAST}, and assume that all photons with optical depths $\tau\geq1$ are absorbed, whereas the rest travel freely~\cite{Mesinger:2010ne}.
This makes the mean-free path of an X-ray photon dependent on its energy. We can, then, define the minimum energy $E_{\rm min}(z,z')$ that a photon can have to travel from $z'$ to $z$, by solving for $\tau=1$, or equivalently, for $\lambda_X(E_{\rm min}) = c (z'-z)/H(z)$, where $\lambda_X$ is the comoving mean-free path (from  Ref.~\cite{Furlanetto:2009uf}), to find
\be
E_{\rm min}(z,z') =  0.3 \,{\rm keV} \left[\left(\frac{1+z'}{15}\right)^2 \dfrac{z'-z}{5 H(z')} \right]^{1/3},
\ee
capping it at $E_{\rm min} = E_0$ if lower.
Here, and throughout, we assume that the spectrum of emission is log-flat in energy, (i.e., $L_X\propto E^{-\alpha_X}$ with $\alpha_X=1$), from a threshold energy  $E_0=0.2$ keV, below which we assume all X-rays are self-absorbed near the sources, up to $E_{\rm max}=2$ keV, above which X-ray photons have mean-free paths large enough to barely heat up the IGM~\cite{Pacucci:2014wwa,Fialkov:2014kta,Greig:2017jdj}.

Under these assumptions, the gas heating is given by an integral of the form~\cite{Mesinger:2010ne}
\be
\epsilon_X (E_{\rm min}) \propto \int_{E_{\rm min}}^{E_{\rm max}} dE   (E - E_i) \sigma_i (E) \left(\dfrac{E}{E_0}\right)^{-\alpha_X-1}
\ee
where $E_i$ and $\sigma_i$  are the energy threshold and cross section for ionization.
For our purposes it suffices to take the large-$E$ behavior of $\sigma_i\propto E^{-7/2}$, to find that the amount of energy deposited into the IGM by X-rays with a minimum energy $E_{\rm min}$ scales as $\epsilon_X\propto E_{\rm min}^{-7/2}$.
We can, therefore, define the ratio $\mathcal R_F$ of the X-ray flux $\mathcal F$ deposited by photons above some energy $E_{\rm min}$, compared to the full spectrum down to $E_0$, as
\be
\mathcal R_F(z,z') = \dfrac{\mathcal F [E_{\rm min}(z,z')]}{\mathcal F(E_0)} =  \dfrac{E_{\rm min}^{-7/2}-E_{\rm max}^{-7/2}}{E_0^{-7/2}-E_{\rm max}^{-7/2}}.
\ee
Then we can write the window function in analogy with the Lyman-$\alpha$ case  simply as
\be
W_X(k,z) = \int\, dz'\,\mathcal R_F(z,z') \dfrac{d F_{\rm coll}}{dz'} \sinc\left[\dfrac{k (z'-z)}{H(z')}\right].
\label{eq:windowX}
\ee

\begin{figure}[hbtp!]
	\includegraphics[width=0.48\textwidth]{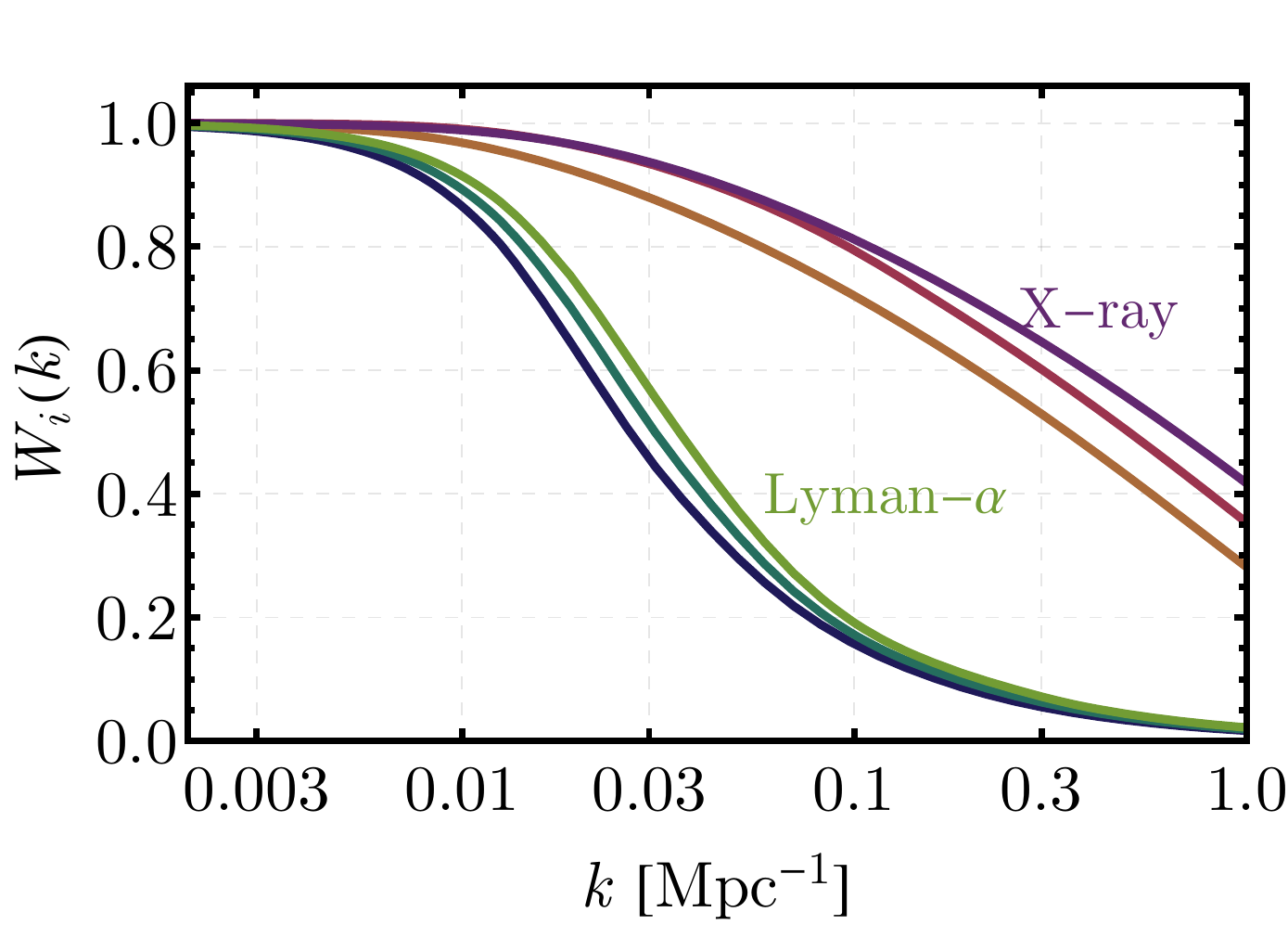}
	\caption{Window functions required to account for photon propagation in the case of X-ray and Lyman-$\alpha$ photons, computed as in Eqs.~(\ref{eq:windowLyA}) and (\ref{eq:windowX}), at redshifts (from left to right)  $z=\{22.3, 25.0, 28.3\}$ for the Lyman-$\alpha$ case and $z=\{14.5, 17.2, 20.3\}$ for the X-ray case.
	The X-ray case shows a less-marked suppression, as X-ray photons typically travel smaller distances than their UV counterparts.
	}
	\label{fig:windowfunc}
\end{figure}

\begin{figure}[hbtp!]
	\includegraphics[width=0.48\textwidth]{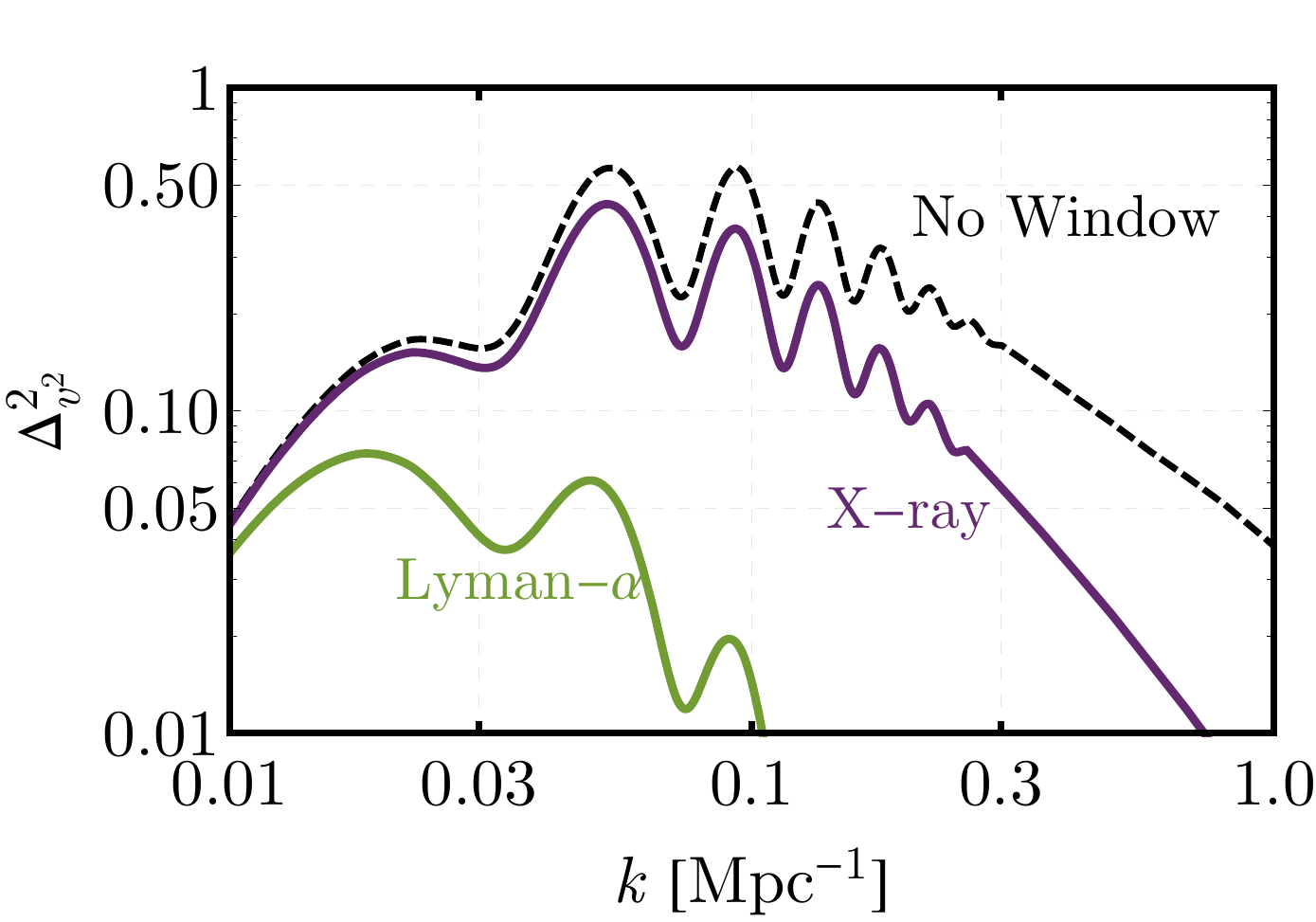}
	\caption{Amplitude of fluctuations of $\delta_{v^2}\equiv \sqrt{3/2} \left [\,(\vcb/v_{\rm rms})^2-1\right ]$ as a function of wavenumber $k$, under three assumptions.
	In the dashed black line we show the local power spectrum (i.e., without any window function), whereas we account for photon propagation by multiplying this quantity by the window function squared for the X-ray and Lyman-$\alpha$ cases in the purple and green lines, corresponding to 21-cm fluctuations at $z=17.2$ and $25.0$, halfway through the EoR and LCE, respectively.
	}
	\label{fig:powvsq}
\end{figure}

Using all these results we show the X-ray window function in Fig.~\ref{fig:windowfunc}, along with that of Lyman-$\alpha$ photons, each at their respective relevant redshifts.
Interestingly, we see that X-ray photon anisotropies are not severely damped within the scales of interest ($k \lesssim 0.3$ Mpc$^{-1}$), where velocity fluctuations are important.
This is in agreement with the X-ray window function obtained for regular overdensities in Ref.~\cite{Fialkov:2015fua}.
We further illustrate this point in Fig.~\ref{fig:powvsq} by showing $\Delta^2_{v^2}$ (the power spectrum of $\delta_{v^2}$) multiplied by window functions (squared) due to either X-ray or Lyman-$\alpha$ propagation, at $z=17.2$ and $z=25.0$ respectively.
We see how accounting for X-ray propagation dampens the small-scale power, while leaving the interesting region of $0.01\leq k\,{\rm Mpc} \leq 0.3$ roughly unaltered, whereas the propagation of Lyman-$\alpha$ photons dampens power at all relevant scales, making the power spectrum difficult to observe beyond $k\gtrsim 0.1$ Mpc$^{-1}$.

\subsection{Size of the Effect}

Now that we understand how photon propagation alters the shape of the $\vcb$ fluctuations, let us fit for the amplitude of the VAOs in our simulations.

As mentioned at the beginning of this section, the VAOs are, to first order, statistically independent from density fluctuations~\cite{Dalal:2010yt}. 
We, therefore, choose to model the 21-cm power spectrum simply
 as
\be
\Delta_{21}^2(k,z) = \Delta_{21,\rm vel}^2(k,z) + \mathcal P_n(k,z),
\label{eq:fitpowsp}
\ee
with the VAOs parametrized as
\be
\Delta_{21,\rm vel}^2(k,z) = A_{\rm vel}(z) \Delta_{v^2}^2(k) \left | W_i(k,z)\right|^2,
\label{eq:delta21vel}
\ee
where $A_{\rm vel}(z)$ is a redshift-dependent amplitude of fluctuations, with units of mK$^2$, which we will fit from the data, and we set the window function to be $W_i(k,z) = W_X(k,z)$ during the EoH, and  $W_\alpha(k,z)$ during the LCE.
The non-VAO component, $\mathcal P_n$, holds a wealth of information on astrophysical parameters~\cite{Mao:2008ug,Mesinger:2013nua,Ewall-Wice:2015uul,DeBoer:2016tnn,Greig:2017jdj,Park:2018ljd}.
Nonetheless, as here we are only interested in the VAOs we will model $\mathcal P_n$ as a simple fourth-order polynomial,
\be
\log \mathcal P_4(k,z) = \sum_{j=0}^4 a_j(z) \left[\log(k)\right]^j,
\ee
which captures the smooth behavior of the power spectrum without wiggles. 
This $\mathcal P_n$ is not designed to provide an excellent fit to the 21-cm power spectrum, but merely to allow us to marginalize over the smooth part of the signal.

We employed the fit from Eq.~\eqref{eq:fitpowsp} in Figs.~\ref{fig:Pow21LyA}, \ref{fig:Pow21Xray}, and \ref{fig:Pow21comparison}, to compare with the simulation results.
In these fits we only used modes with $k\geq0.05$ Mpc$^{-1}$, to avoid the sizable error bars of large-scale modes in the boxes.
We see that in all cases our model provides a reasonable fit for the 21-cm fluctuations over the entire relevant range where velocities are important ($0.03\,\rm Mpc^{-1} \leq {\it k} \leq 0.3 \, Mpc^{-1} $).
For reference, at the redshifts that we showed in Figs.~\ref{fig:Pow21LyA} and \ref{fig:Pow21Xray} the value of the VAO amplitude is $A_{\rm vel} = \{330, 1100, 36, 270, 27, 0.02 \}$ mK$^2$, for decreasing redshift.
This amplitude always increases the large-scale 21-cm power spectrum, allowing for an easier detection of this signal.

\begin{figure}[hbtp]
	\includegraphics[width=0.48\textwidth]{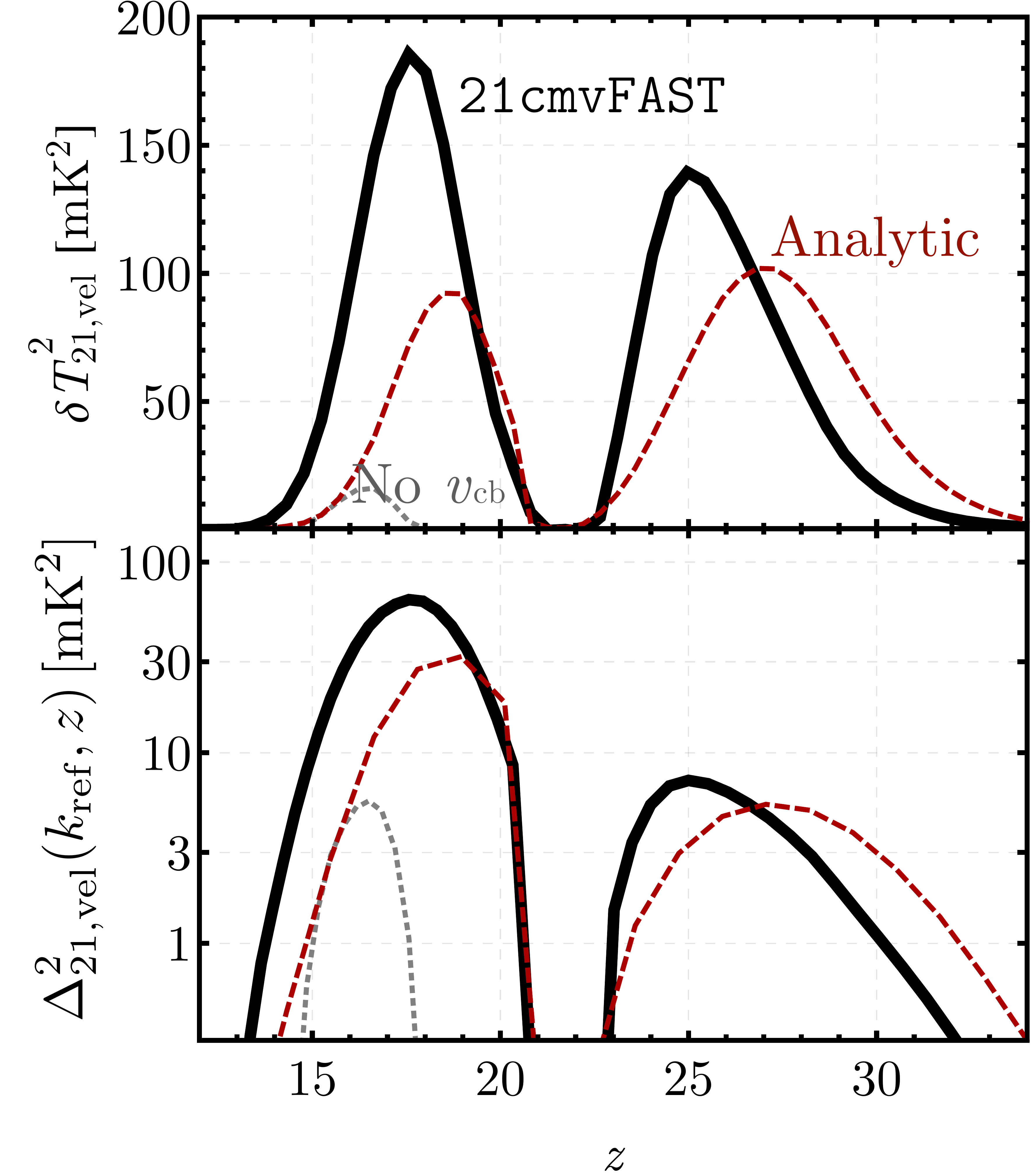}
	\caption{Amplitude of the 21-cm VAOs as a function of redshift.
		In the top panel we show the VAO-induced variance, computed from Eq.~\eqref{eq:21variance}, obtained from our simulations in the black line and from analytic estimates in the red dashed line. 
		As a consistency check we show the result from a simulation without $\vcb$ fluctuations in the grey dotted line, which should be consistent with zero, providing an estimate of the error in our fitting procedure.
		In the bottom panel we show the VAO-only power spectrum for the same case, at a reference scale $k_{\rm ref}=0.14 \, \rm Mpc^{-1}$, typically observable by interferometers.
	}
	\label{fig:amplitudes}
\end{figure}

This $A_{\rm vel}$ is, however, not directly observable, as it depends on the assumptions about the shape of the VAOs (i.e., the window functions).
In order to better quantify the size of the VAOs we define the variance in  $T_{21}$ induced by the velocities alone as
\be
\delta T^2_{21,\rm vel} (z) = \int d\log k\, \Delta_{21,\rm vel}^2(k,z),
\label{eq:21variance}
\ee
with $\Delta_{21,\rm vel}^2$ as defined in Eq.~\eqref{eq:delta21vel}, which depends on $A_{\rm vel}$ and takes into account the suppression in fluctuations due to photon propagation.
We show this VAO-sourced variance as a function of redshift in Fig.~\ref{fig:amplitudes}, where we can distinguish the two peaks due to the  Lyman-$\alpha$ coupling and X-ray heating, the former peaking at $z\approx24$ and the latter at $z\approx18$.
Additionally, we reach the expected result that at high redshifts there are not enough UV photons for efficient Lyman-$\alpha$ coupling (so the VAO amplitude all but vanishes above $z\gtrsim 30$), whereas at small redshifts ($z\lesssim 14$) the gas is heated up above $T_\gamma$, both of which render the effect of VAOs negligible.
Interestingly, the VAO amplitude cancels in the transition from the LCE to the EoH.
This is because the effect of 
velocities on the 21-cm line during the LCE and the EoH goes in opposite directions.
During the LCE a large $\vcb$ results in higher $T_{21}$, as fewer Lyman-$\alpha$ photons couple $T_s$ to $T_g$.
On the other hand, during the EoH larger velocities produce a more-negative $T_{21}$, as fewer X-ray photons heat up the gas.
This means that, by necessity, these two components have to cancel at some point, resulting in nearly zero VAO fluctuations, which in our simulations occurs at $z \approx 21$.

In addition to the VAO-sourced 21-cm variance $\delta T_{21,\rm vel}^2$, we show in Fig.~\ref{fig:amplitudes} the VAO-only power spectrum $\Delta_{21,\rm vel}^2$ as a function of redshift at a reference scale $k_{\rm ref}=0.14\,\rm Mpc^{-1}$, computed with Eq.~\eqref{eq:delta21vel}.
Here the effect of the window functions is obvious, as the fluctuations on $T_{21}$ during the LCE are significantly smaller than during the EoH at this wavenumber, even though the effect of the velocities in $F_{\rm coll}$ is larger at higher redshifts.
As a test, we also show in Fig.~\ref{fig:amplitudes} results for a case in which we fix $\vcb$ to its average value, obtained through the same fitting procedure as our main result. 
This curve can be taken as an estimate of the errors in both the fitting procedure and our simulations, and is significantly smaller than the amplitude in the case with velocities.
We therefore conclude that our procedure is properly capturing both the shape and amplitude of the VAOs.

\subsection{Comparison with Analytic Calculations}

We now estimate the expected size of the VAO signal from analytic arguments, extending on the work of Refs.~\cite{Dalal:2010yt,McQuinn:2012rt} by including X-ray and Lyman-$\alpha$ fluctuations jointly.
During cosmic dawn we can safely neglect collisional hyperfine transitions, so the spin temperature will be given by~\cite{Pritchard:2011xb}
\be
T_s^{-1} = \dfrac{T_\gamma^{-1} + x_\alpha T_{g}^{-1}}{1+x_\alpha},
\ee
where $x_\alpha\propto n_\alpha$ accounts for the strength of Lyman-$\alpha$ pumping (see Eq.~\eqref{eq:nalpha}).
This allows us to recast Eq.~\eqref{eq:T21} and write the fluctuations in the 21-cm brightness temperature as
\be
\delta T_{21} = 38\,{\rm mK }\, \left(\dfrac{1+z}{20}\right)^{1/2} \dfrac{1 + x_\alpha T_\gamma/T_{g}}{1+x_\alpha}  \kappa(z) ,
\label{eq:deltaT21vel}
\ee
where we have ignored density fluctuations in this simplistic model, and $\kappa(z)$ is a correction factor that we will describe below.

As illustrated above, the DM-baryon relative velocities modulate the collapse fraction $F_{\rm coll}$, which in turn alters the 21-cm temperature through both the $x_\alpha$ and $T_{g}$ terms.
In this section we will follow our simple model in which the amount of X-ray and Lyman-$\alpha$ photons are both proportional to $F_{\rm coll}$, so that
\begin{subequations}
\ba
x_\alpha(\vcb) &= \VEV{x_\alpha} \dfrac{F_{\rm coll}(\vcb)}{\VEV{F_{\rm coll}}}, \\
T_{g}(\vcb) &= T_{g}^{(\rm ad)} + \VEV{\Delta T_X} \dfrac{F_{\rm coll}(\vcb)}{\VEV{F_{\rm coll}}},
\end{align}
\end{subequations}
where  $T_g^{(\rm ad)}$ is the gas temperature considering only adiabatic cooling, and $\Delta T_X$ is the change in gas temperature due to X-ray heating.
For practical purposes we use the box-averaged values of $x_\alpha$ and $\Delta T_X$ from our simulations, instead of computing them from first principles.
Additionally, given that we are ignoring density fluctuations in this calculation, we will not obtain the correct background evolution.
We thus include a correction factor $\kappa(z) = \VEV{T_{21}}_{\rm sim}/\VEV{T_{21}}_{\rm ana}$ in Eq.~\eqref{eq:deltaT21vel}, in order to recover the correct average 21-cm signal.

With our toy model in Eq.~\eqref{eq:deltaT21vel}, we can calculate the amplitude of the VAOs using the result of Refs.~\cite{Dalal:2010yt,Ali-Haimoud:2013hpa,Munoz:2018jwq}---here recast as in Eq.~\eqref{eq:powspf}---and defining the 21-cm velocity bias as
\be
b_{21}^2 = \VEV{T_{21}^2(\vcb)} - \VEV{T_{21}(\vcb)}^2.
\ee
Then, the VAO component of the power spectrum will be
\be
\Delta_{21,\rm vel}^2(k,z) = b_{21}^2(z) \Delta^2_{v^2}(k) \left | W_i(k,z)\right|^2,
\ee
which has the theoretically motivated shape, and an amplitude given by $b_{21}^2$.
We compare the results for this analytic calculation with those of our simulations in Fig.~\ref{fig:amplitudes}, where we can see that the main features of the signal are well reproduced by this approximation.
For instance, we see the rise in power from early times when the Lyman-$\alpha$ pumping turns on, peaking at $z\approx25$, and the cancellation between the Lyman-$\alpha$ and X-ray terms at $z\approx 21$,
followed by the increase during the X-ray heating phase, at $z\approx18$.

\section{Total signal and observability}
\label{sec:obs}

The inclusion of streaming velocities produces VAOs in the 21-cm power spectrum on large scales.
We now explore how the signal changes when changing different assumptions, especially the poorly known Lyman-Werner feedback, as well as how observable these VAOs are by upcoming 21-cm interferometers.

\subsection{Different Feedback Assumptions}

We start by considering the VAOs under three additional feedback assumptions, different than the regular-feedback strength that we have focused on so far.
The first is an unrealistic no-feedback case, which will show the largest 21-cm VAOs.
The other two are realistic feedback strengths: low and high, slightly more optimistic and pessimistic than our fiducial case, as explained in Section~\ref{sec:vcbeffect}.
We also explore the 21-cm power spectrum for the unphysical case that LW feedback is saturated at all times (so only atomic-cooling haloes can form stars) in App.~\ref{App:atomic}.

\begin{figure}[hbtp]
	\includegraphics[width=0.5\textwidth]{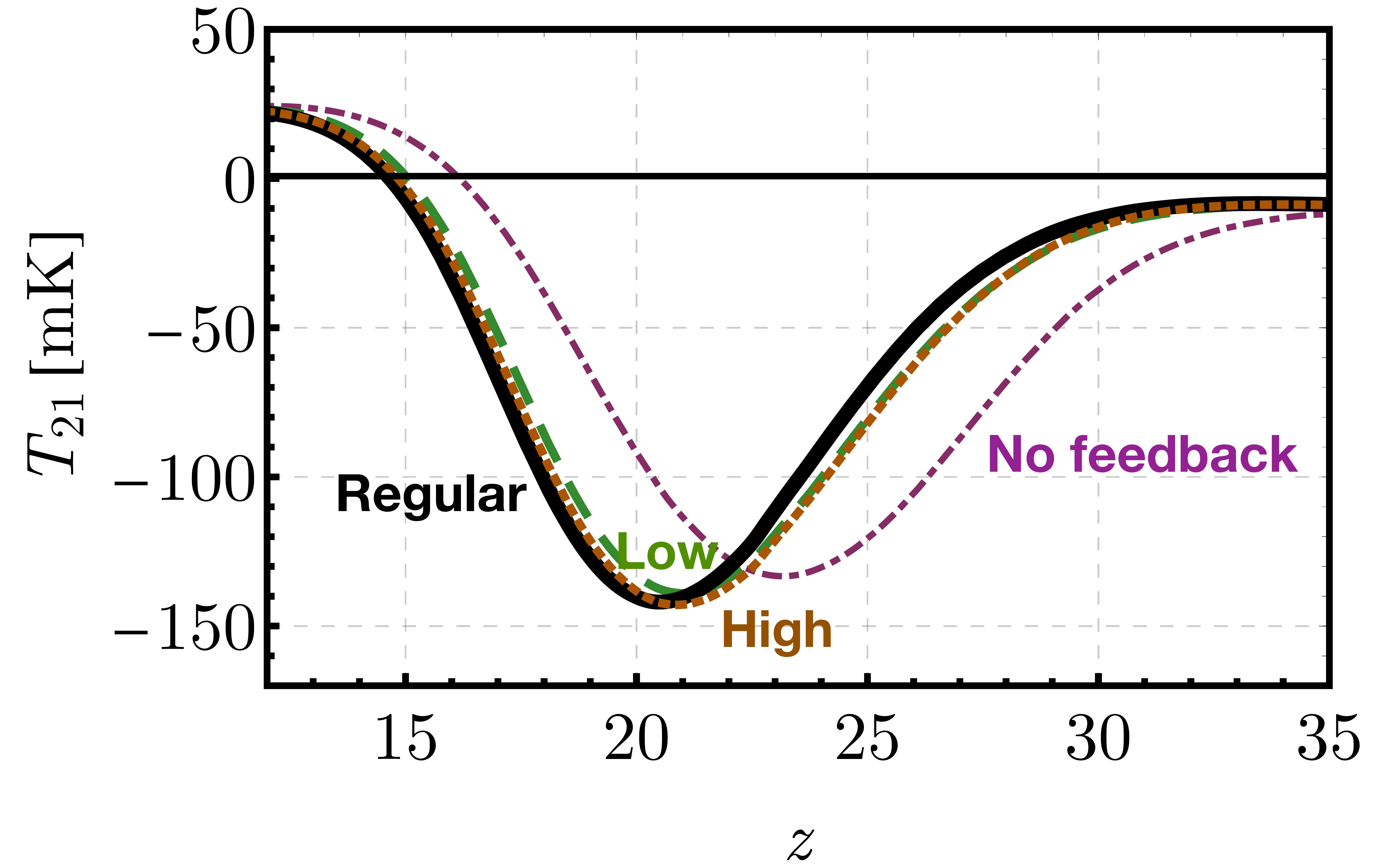}
	\caption{Global 21-cm signal for the four feedback cases we study. From right to left we show the cases of no feedback, as well as low-, high-, and regular-feedback strengths, in all cases including the effect of $\vcb$.	
	}
	\label{fig:T21avgFDBCK}
\end{figure}

We first show, in Fig.~\ref{fig:T21avgFDBCK}, the evolution of the global 21-cm signal as a function of redshift for the four feedback cases. Here, barring the effect of the LW feedback, we have kept fixed all the fiducial parameters (except in the no-feedback case, where we have a lower $f_*=0.03$).
Adding feedback always delays evolution, as argued in Section~\ref{sec:21cmline}, which makes the three realistic feedback cases  delayed by $\Delta z\sim 3$ with respect to the no-feedback case.
Within the three realistic cases, however, the low feedback is the fastest, due to the smaller effect of LW radiation, followed by the high-feedback case, which is affected by LW as much as the regular case, while the $\vcb$ effect is smaller (due to the assumption of uncorrelated feedback), producing smaller halo suppression.
The regular case is the most delayed.

\begin{figure}[hbtp]
	\includegraphics[width=0.5\textwidth]{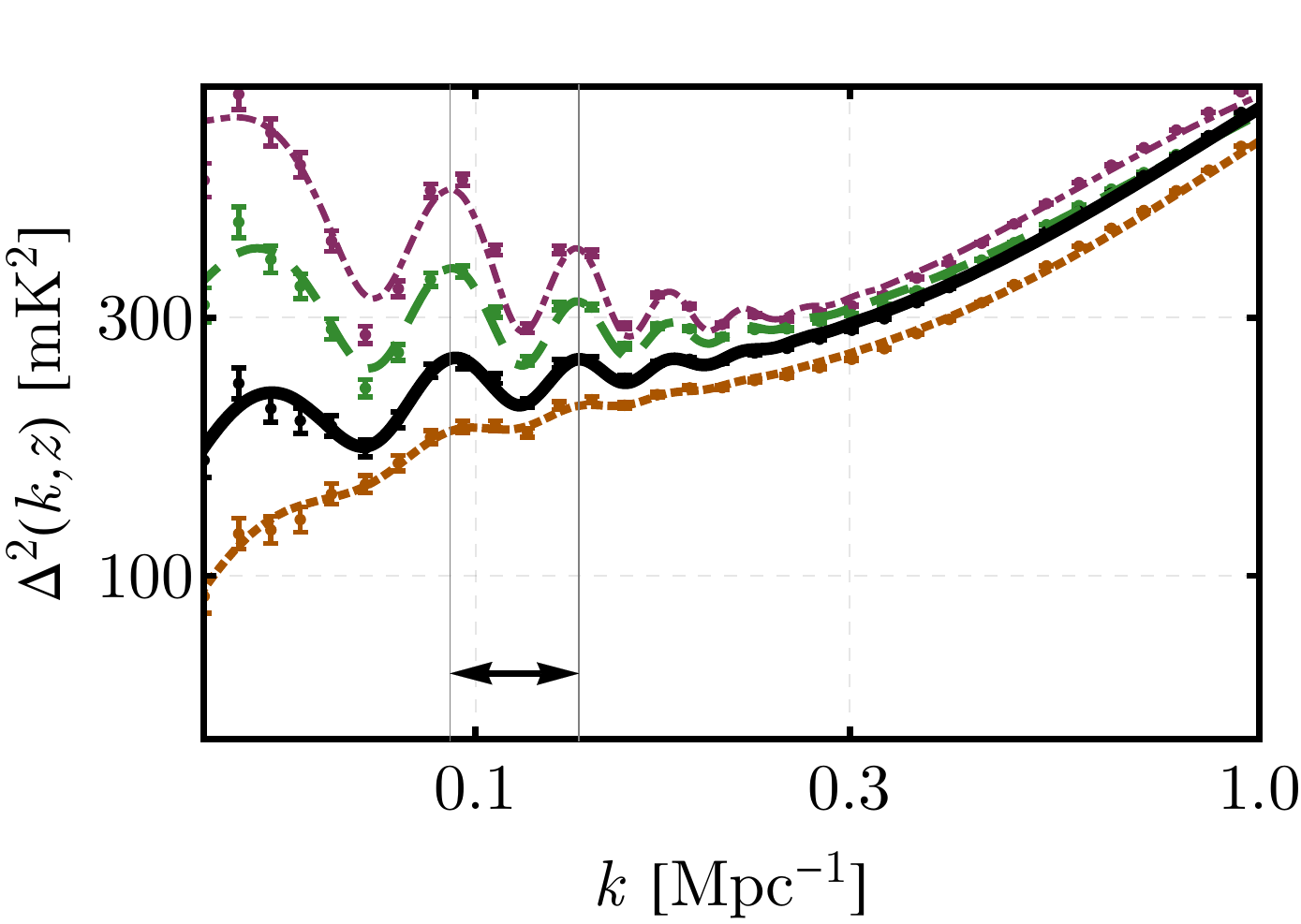}	
	\caption{The 21-cm power spectrum halfway through the epoch of heating (where  $\VEV{T_{21}} = 0.5\times \VEV{T_{21}}_{\rm min}$), for each feedback case. From top to bottom we show the cases of no feedback, and low-, regular-, and high-feedback strengths, with corresponding redshifts  $z=\{ 19.1, 17.6, 17.2, 17.2\}$.	
		The vertical gray lines are separated by $\Delta k=2\pi/r_{\rm drag}$, where $r_{\rm drag}=147$ Mpc is the acoustic scale, as the shape and periodicity of the VAOs do not depend on the feedback strength.
	}
	\label{fig:P21FDBCKXray}
\end{figure}

\begin{figure}[hbtp]
	\includegraphics[width=0.5\textwidth]{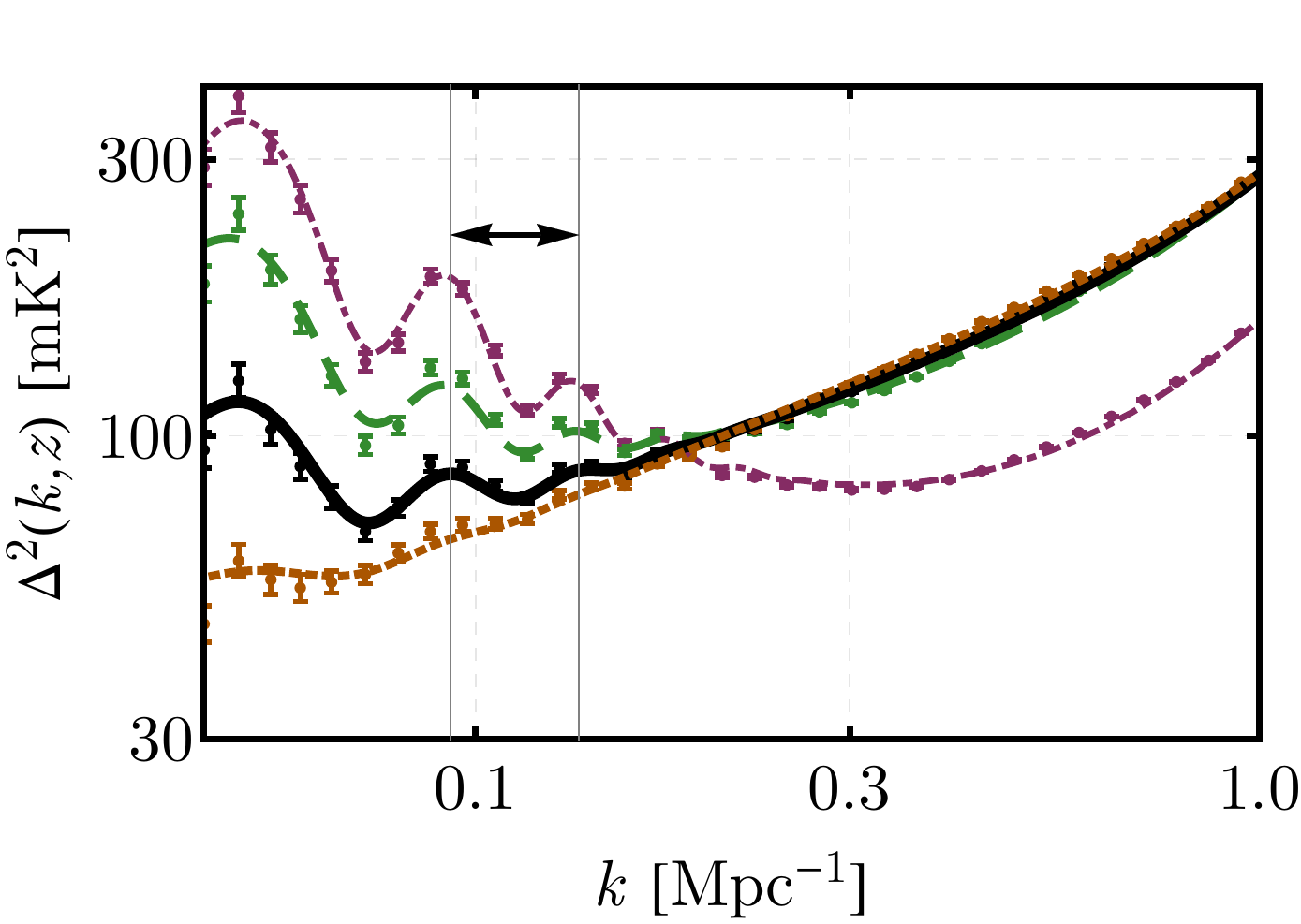}	
	\caption{Same as Fig.~\ref{fig:P21FDBCKXray} but for the halfpoint of the Lyman-$\alpha$ coupling era, with $\VEV{T_{21}} = 0.5\times \VEV{T_{21}}_{\rm min}$, which corresponds to $z=\{28.3, 25.5, 25.0, 25.5\}$, going from lowest to highest feedback strength.
	}
	\label{fig:P21FDBCKLyA}
\end{figure}

Even though the global signal does not vary significantly between the different feedback cases, the large-scale 21-cm fluctuations will.
We show the 21-cm power spectra for all feedback cases at the midpoint of the EoH in Fig.~\ref{fig:P21FDBCKXray} (corresponding to $z=\{19.1, 17.6, 17.2, 17.2\}$, in ascending order of feedback strength), and at the midpoint of the LCE in Fig.~\ref{fig:P21FDBCKLyA} ($z=\{27.7, 25.5, 24.5, 25.5\}$)\footnote{We note that {\tt 21cmFAST} has a redshift resolution of $\Delta z \approx 0.3$ during cosmic dawn, which determines our precision on all redshifts reported.}.
In all cases we see a similar small-scale $k\gtrsim 0.3\,\rm Mpc^{-1}$) power at a given redshift, while increasing the feedback lowers the large-scale power due to the VAOs.
This is expected, since stronger feedback will prevent smaller-mass molecular-cooling haloes to form stars, thus rendering some of the suppression of the relative velocities ineffective.
In particular, we see in Fig.~\ref{fig:P21FDBCKXray} how the no-feedback case gives rise to large ($\Delta_{21}^2 \gtrsim 400$ mK$^2$) acoustic power at low-$k$, whereas that power is reduced to $\Delta_{21}^2\approx 200$ mK$^2$ for the realistic feedback cases.
The situation is similar during the LCE, albeit with smaller power overall.
Interestingly, in all cases our curves provide a good fit (through Eq.~\eqref{eq:fitpowsp}) to the simulation data at relevant scales ($k = 0.03-0.3$ Mpc$^{-1}$), showing that the effect of astrophysical feedback is to lower the amplitude of the VAO signal, but not to alter its shape, which is indeed that of $\Delta^2_{v^2}$.

\begin{figure}[hbtp!]
	\includegraphics[width=0.5\textwidth]{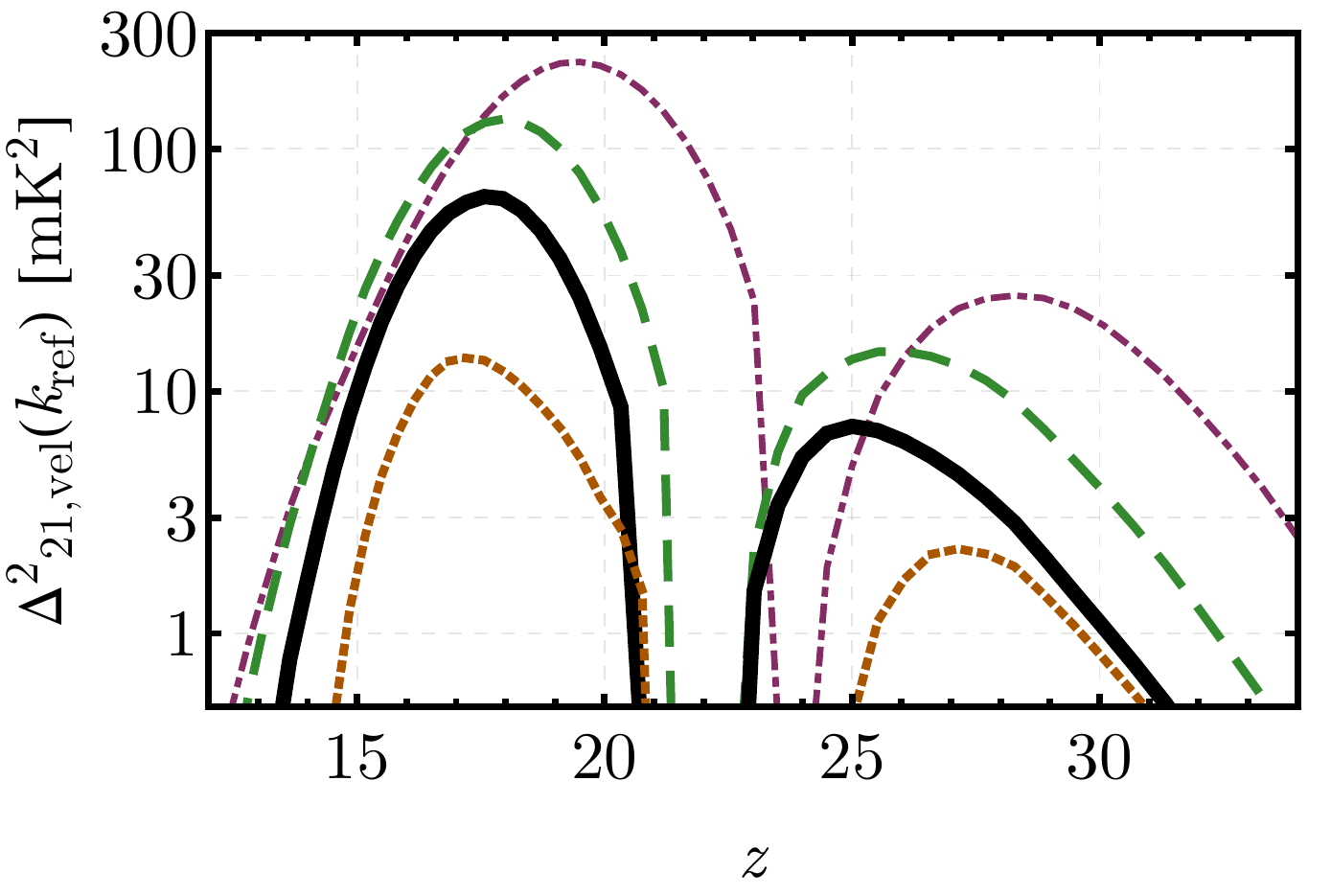}	
	\caption{Same as the bottom panel of Fig.~\ref{fig:amplitudes}, albeit for the four feedback cases, from top to bottom: no feedback, low, regular, and high feedback.
	Higher feedback produces a smaller VAO amplitude, as stellar production in molecular-cooling haloes is less important.
	}
	\label{fig:amplitudesFDBCK}
\end{figure}

Finally, we show the amplitude of the VAOs as a function of redshift in Fig.~\ref{fig:amplitudesFDBCK} for the same four feedback cases.
All the curves follow a similar trend, with the VAO amplitude rising at $z \sim 25$ due to Lyman-$\alpha$ pumping, and again at $z \sim 18$ due to X-ray heating, with the transition in between having nearly vanishing velocity-induced power.
Nonetheless, the size of the VAOs changes dramatically for different feedback assumptions. 
Focusing on the peak of each curve during the EoH, the power spectrum at a reference scale $k_{\rm ref}=0.14\,\rm Mpc^{-1}$ is reduced from 234 mK$^2$ in the no-feedback case, to 138 mK$^2$ for low feedback strength, and even further to 14 mK$^2$ in the pessimistic case of high feedback.
In the regular case studied in the previous section this amplitude is 64 mK$^2$.
This large range brackets the possible amplitude of the VAOs during cosmic dawn, depending on the thus far unknown physics of the first stellar formation.
Nonetheless, in all feedback cases, even with pessimistic assumptions, the power spectrum shows nonzero $\vcb$-induced 21-cm perturbations.
Our results are in agreement with the overall trends in Ref.~\cite{Fialkov:2013jxh}, albeit with some distinctions given the different choices of astrophysical and feedback parameters, which we will discuss below.

\subsection{Observability}

While the 21-cm signal due to the DM-baryon relative velocities is significant during both the EoH and the LCE, the large foregrounds make the task of observing---and characterizing---it challenging.
We now forecast the sensitivity of large 21-cm observatories to the VAO signal, focusing on the upcoming Hydrogen Epoch of Reionization Array (HERA)\footnote{\url{ https://reionization.org} }~\cite{DeBoer:2016tnn},
for which we obtain the projected power-spectrum uncertainties with the publicly available {\tt 21cmSense}~\cite{Pober2013,Pober:2013jna}\footnote{\url{ https://github.com/jpober/21cmSense } }.
This code accounts for the $u$-$v$ sensitivities of each antenna in an array, and returns the errors in the 21-cm power spectrum, including cosmic variance.

In addition to the instrumental noise of the antennae, one of the main obstacles to any prospective 21-cm detection is the presence of large foregrounds, from  both our Galaxy and the atmosphere, that contaminate the signal.
These foregrounds swamp any primordial 21-cm signal in the majority of the Fourier plane, as wavenumbers with
\be
k_{||} \leq a + b k_\perp
\ee
are part of the so-called ``foreground wedge", and not expected to be usable for cosmology~\cite{Liu:2009qga,Datta:2010pk,Morales:2012kf,Parsons:2011ew,Parsons:2012qh}. 
Here $k_{||}$ and $k_\perp$ are the line-of-sight (LoS) and perpendicular wavenumbers, and $a$ and $b$ parametrize the severity of the foregrounds by accounting for how many wavenumbers are discarded.
We follow the parametrization in {\tt 21cmSense}, and define three foreground scenarios:

$\bullet$ Pessimistic, with $a=0.1\,h\,\rm Mpc^{-1}$, and $b$ given by the horizon limit, as usual.

$\bullet$ Moderate, with $a=0.05\,h\,\rm Mpc^{-1}$, and the same $b$.

$\bullet$ Optimistic, with $a=0$, and a smaller $b$ by a factor of $\sin(\theta_{\rm FWHM}/2)$, where $\theta_{\rm FWHM}$ is the full-width half-maximum of the beam.

Note that here we report wavenumbers multiplied by the reduced Hubble constant $h$, as those are the inputs of {\tt 21cmSense}.
In all cases we assume that the sensitivity of different baselines can be added coherently, and we take a system temperature of $T_{\rm sys} = 100 + 120 (\nu/150\,\rm MHz)^{-2.55} \,\rm K$, as in Ref.~\cite{DeBoer:2016tnn}.
For reference, the wedge has been experimentally observed in Ref.~\cite{Pober:2013ig}, with $a$ and $b$ similar to the pessimistic case, although it is likely that a better understanding of the foregrounds and the instrument will improve these parameters.

We divide the observation range into four redshift bins, three of them in the EoH, centered at redshifts $z=14,16,$ and 18, each with a width $\Delta z=2$, and one in the LCE, centered at $z=25$ and with $\Delta z=4$.
We use the HERA Split Core configuration, with 540 days of observation (corresponding to roughly three full years), and the standard settings of {\tt 21cmSense}, albeit binning logarithmically in $k$ instead of linearly to better capture the VAO shape.

We calculate the signal-to-noise ratio (SNR) from each redshift slice summing simply over $k-$bins as
\be
{\rm SNR}^2 (z) = \sum_{k\rm -bins} \left[
\dfrac{ \Delta^2_{21,\rm vel}(k,z)}{\delta(\Delta^2_{21})} \right]^2,
\ee
where the binned errors $\delta(\Delta^2_{21})$ in the 21-cm power spectrum  are the output of {\tt 21cmSense}, and we consider the range $k=(0.05-0.5)\, h\,\rm Mpc^{-1}$.
Note that we are only computing the SNR of the VAOs, although, of course, we include the full signal in $\delta(\Delta^2_{21})$ to properly account for cosmic variance.
We show our results in Table~\ref{tab:SNRs} separately for the EoH and the LCE, considering each combination of foreground and feedback cases studied in this paper, where the SNRs of the EoH bins have been added in quadrature.
We see that HERA has a high likelihood of observing the VAOs during the EoH, where our regular-feedback case yields SNRs of 10-62, and even the most pessimistic feedback assumptions predict a SNR larger than unity.
The situation is significantly worse during the LCE, where only under highly optimistic assumptions will it be possible to detect the VAO signal at SNR larger than 5.
The cases with the largest SNRs yield the most promising cosmological measurements from VAOs~\cite{PaperII}.

\begin{table}[hbtp!]
	\noindent\begin{tabular}{ c | c  c  c }	
		\multicolumn{1}{c}{} 	&\multicolumn{3}{c}{Foregrounds}\\
		\cline{2-4}
		Feedback strength  &  Pessimistic & Moderate &  Optimistic \\
		\hline
		High &  2  &   4   & 14  \\
		
		Regular  &  10  & 15  &   62   \\
		
		Low  &  20  & 33  &  133   \\
		
	\end{tabular}
	\noindent\begin{tabular}{ c | c  c  c }	
	\multicolumn{1}{c}{} 	&\multicolumn{3}{c}{Foregrounds}\\
	\cline{2-4}
	Feedback strength  &  Pessimistic & Moderate &  Optimistic \\
	\hline
	High & $<1$  &  $<1$ &  $<1$   \\
	
	Regular  & $<1$  &  3  &  29  \\
	
	Low  &  2   &  6   &  52   \\
	
	\end{tabular}
	\caption{Projected signal-to-noise ratios for each combination of feedback  and foreground assumptions, in the first table for the X-ray heating era ($z=13-19$) and in the second table for the Lyman-$\alpha$ coupling era ($z=23-27$).
	In all cases we assume three years of observation with HERA.
	}
	\label{tab:SNRs}
\end{table}

As a test, we have run simulations varying our fiducial X-ray luminosity by a factor of six upwards (so $\log_{10}(L_X/{\rm SFR})=40.8$), to determine how much this parameter affects the observability of the VAOs.
We show our results for this case in Table~\ref{tab:SNRs2}.
On the one hand, increasing the luminosity shifts all heating transitions earlier (for this case $\Delta z \approx 2$ away from our fiducial case, as for instance we find $z_0=17.7$), where feedback effects were less severe, thus increasing the (relative) size of the VAOs.
On the other hand, the overall 21-cm absorption is shallower, as the gas is warmer by the time Lyman-$\alpha$ coupling is completed.
We find that the velocity-induced power spectrum at its redshift peak $z=19.7$ (and scale $k_{\rm ref}=0.14\,\rm Mpc^{-1}$) is $\Delta_{\rm vel}^2(k_{\rm ref})=65$ mK$^2$. 
This is very similar to our fiducial case, showing that
the effect of weaker feedback at large redshifts is compensated by the shallower 21-cm absorption, which sets the overall scale of the signal.
There is one more difference between these two cases, however. 
Since all heating transitions occur earlier for higher X-ray luminosity,
an increased noise (due to the higher system temperature) will be present in these large-$z$ observations, which lowers the SNRs for this case in Table~\ref{tab:SNRs2}.

\begin{table}[hbtp!]
	\noindent\begin{tabular}{ c | c  c  c }	
		\multicolumn{1}{c}{} 	&\multicolumn{3}{c}{Foregrounds}\\
		\cline{2-4}
		Feedback strength  &  Pessimistic & Moderate &  Optimistic \\
		\hline
		High &  2  &   4   & 23  \\
		
		Regular  &  7  & 11  &   47   \\
		
		Low  &  12  & 20  &  83   \\
		
	\end{tabular}
	\noindent\begin{tabular}{ c | c  c  c }	
		\multicolumn{1}{c}{} 	&\multicolumn{3}{c}{Foregrounds}\\
		\cline{2-4}
		Feedback strength  &  Pessimistic & Moderate &  Optimistic \\
		\hline
		High & $<1$  &  $<1$ &  6  \\
		
		Regular  & $<1$  &  2  &  17  \\
		
		Low  & $<1$   &  3   &  19   \\
		
	\end{tabular}
	\caption{Same as Table~\ref{tab:SNRs}, albeit with an X-ray luminosity of $\log_{10}(L_X/{\rm SFR})=40.8$, a factor of 6.3 larger than our fiducial case.
	This pushes all heating transitions $\Delta z\sim 2$ earlier, so the EoH covers the range $z=15-21$.
	}
	\label{tab:SNRs2}
\end{table}

\subsection{Comparison to Previous Work}

We are not the first to compute the effect of the DM-baryon streaming velocities in the 21-cm power spectrum, although we are the first to do so with both a (public) simulation code and analytical results.
This has allowed us to confirm that the well-understood shape of the VAOs is indeed present in 21-cm simulations.
We will now compare our results with those of previous work on this topic.

On the theory side, the authors in Ref.~\cite{Dalal:2010yt} were the first to predict and calculate the velocity-induced 21-cm fluctuations due to Lyman-$\alpha$ pumping.
In that work it was predicted that the acoustic shape of fluctuations would follow $\Delta^2_{v^2}$ with suppression on small scales, exactly as we found in our simulations, where the window function posited by Ref.~\cite{Dalal:2010yt} provides a good fit to our data.
Additionally, that work showed a $\vcb$-induced 21-cm amplitude of $\delta T_{21} \approx $ 3 mK, in line with (albeit smaller than) our result of $\delta T_{21} \approx $ 10 mK at our peak ($z=25$).
Additionally, in Ref.~\cite{McQuinn:2012rt} the VAOs during the EoH were analytically estimated, and were found to produce a 21-cm power spectrum of $\Delta^2_{21} \approx 100$ and 300 mK$^2$ at $z=15$ and $20$, respectively.
This is in line with the results of our simulations.
Note, however, that in Refs.~\cite{Dalal:2010yt,McQuinn:2012rt} the non-VAO part of $\Delta_{21}^2$, which we parametrized as a smooth polynomial $\mathcal P_n$, was assumed to be proportional to the density power spectrum.
This is an overly simplistic assumption, which does not resemble the real 21-cm signal, as is clear from the results of Refs.~\cite{Mesinger:2010ne,Visbal:2012aw,Fialkov:2012su}, or our own Fig.~\ref{fig:Pow21comparison}.
The non-VAO part of the power spectrum behaves as a shallower function of wavenumber $k$, resulting in observable VAOs even at $k\sim0.1$ Mpc$^{-1}$.

On the simulation side, in Ref.~\cite{Visbal:2012aw} the effect of velocities in the 21-cm signal was first implemented, albeit only applied to the X-ray heating era. 
In that study it was found that the relative velocities can produce 21-cm fluctuations that dominate the signal on large scales, and have sharp BAO-like features, as we confirm here.
Their results are comparable to our ``no-feedback" case (purple line in Fig.~\ref{fig:P21FDBCKXray}), which, however, we argue is unlikely to represent reality.
This code was improved upon in Ref.~\cite{Fialkov:2012su}, where the effects of Lyman-Werner feedback were included.
In this case we find broad agreement between their ``strong-feedback" and our "regular-feedback" models, as both have the same feedback parameters, although we find larger-amplitude oscillations, likely because of our homogeneous LW feedback.
Finally, in Ref.~\cite{Fialkov:2013jxh} that work was extended to include fluctuations during the LCE, where the shape of the VAOs seems to be damped at $k\gtrsim 0.1$ Mpc$^{-1}$, which we find here as well.
We note two important differences between our work (based on {\tt 21cmvFAST}) and that of Refs.~\cite{Visbal:2012aw,Fialkov:2012su}.
First, as in {\tt 21cmFAST} we use the EPS formalism  to compute the amount of photons emitted from galaxies some distance $R$ away by  averaging densities and velocities (as described in Section~\ref{sec:vcbeffect}), which in Ref.~\cite{Zahn:2010yw} was found to agree well with full simulations), whereas in Refs.~\cite{Visbal:2012aw,Fialkov:2012su} the collapsed fraction is calculated at each point and then averaged over $R$.
Secondly, for computational simplicity we are only taking the average value of the Lyman-Werner flux across the whole simulation box, as opposed to computing it at every pixel, as done in Ref.~\cite{Fialkov:2012su}.

We note that full hydrodynamical simulations of stellar formation, such as those of Refs.~\cite{Trac:2009bt,Gnedin:2014uta,Kaurov:2015zza,Ocvirk:2015xzu,Poole:2015mhx,Mesinger:2016ddl,Kern:2017ccn}, are expected to be a closer match to reality than the quasi-numerical simulations done here (based on Refs.~\cite{Mesinger:2010ne,Greig:2015qca}) or in Refs.~\cite{Visbal:2012aw,Fialkov:2012su,Fialkov:2013jxh,Fialkov:2014wka,Cohen:2015qta,Cohen:2016jbh}.
Nonetheless, the effects that we are studying chiefly affect haloes below the atomic-cooling threshold, with masses $M \lesssim 10^7\,M_\odot$, which are not resolved in current simulations~\cite{Gnedin:2014uta}.
A significant improvement in the mass resolution is required to resolve the haloes that are most suppressed by the streaming velocities, and we hope that this work stimulates the study of VAOs in those simulations.

\section{Conclusions}
\label{sec:conclusions}

In this work we have studied the effect of the dark matter (DM)-baryon relative velocities on the 21-cm line.
These ``streaming" velocities, sourced by the baryon acoustic oscillations (BAOs), become supersonic after recombination, and suppress stellar formation in molecular-cooling haloes (with masses $M\lesssim {\rm few} \times 10^7\,M_\odot$), which are expected to host the first stars during cosmic dawn.
We performed quasi-numerical simulations with the {\tt 21cmvFAST} code, which we make public.
This allows us to properly include molecular-cooling haloes, as setting $T_{\rm cool} < 10^4$ K in the regular {\tt 21cmFAST} will result in a severe overestimation of the amount of star-forming haloes.

The fluctuations in relative velocities become imprinted onto the distribution of the first stars and, through their X-ray and UV emission, onto the 21-cm line.
This generates velocity-induced acoustic oscillations (VAOs) in the 21-cm power spectrum on large scales, turning the cosmic dawn into a novel probe of acoustic physics, as the BAOs are imprinted onto the DM-baryon relative velocities on their genesis. 
We emphasize that the presence of these acoustic oscillations on the 21-cm line has  previously been pointed out in Refs.~\cite{Dalal:2010yt,McQuinn:2012rt,Visbal:2012aw,Fialkov:2012su,Fialkov:2013jxh}.
Here we have compared, for the first time,
analytic and simulation-based results, in order to isolate the shape and size of the VAOs.
We have found that, to a good approximation, the VAOs follow a well-understood analytic shape, regardless of astrophysical effects such as photon propagation.
We exploit this result in Ref.~\cite{PaperII}, where we propose using VAOs as a standard ruler at cosmic dawn.
Interestingly, here we found that VAOs are observable with the upcoming HERA interferometer under a broad range of astrophysical assumptions.
This would illuminate the physics of the cosmic-dawn era, showing that molecular-cooling haloes play an important role in stellar formation.
In addition, the inclusion of streaming velocities can result in sizable non-Gaussianities, which we plan to explore in future work.

We note that the VAOs we study here are different from the ``regular" (density) BAOs, as they do not originate from small wiggles in over/under-densities, but instead from $\mathcal O(1)$ fluctuations in the relative velocity.
The density BAOs would also affect the 21-cm signal during different cosmic eras~\cite{Barkana:2005xu,Barkana:2005nr,Chang:2007xk,Wyithe:2007rq,Barkana:2010zq}.
Nonetheless, density fluctuations affect the cosmic-dawn 21-cm line in a complicated fashion, hampering a precise reconstruction of the matter power spectrum, so we have not included them in the smooth $\mathcal P_n$ function in Eq.~\eqref{eq:fitpowsp}.
Additionally, while we have included the effect of Lyman-Werner feedback, which increases the minimum mass required to form stars by dissociating molecular hydrogen, there may be other feedback processes, such as chemical~\cite{Springel:2002ux} or other radiative feedback~\cite{Barkana:1999apa,Ahn:2006qu,Mesinger:2008ze,Okamoto:2008sn,Ahn:2008gh} that may increase the halo mass beyond even atomic-cooling haloes, rendering the effects of velocities negligible.
Our simulations show, however, that if the VAO signal is present, it will always have the acoustic shape that we have modeled here.

To summarize, the cosmic-dawn era will provide us with a new arena to search for acoustic oscillations, in the form of VAOs. 
Their detection is within reach of upcoming 21-cm experiments, and will teach us a great deal about the haloes that hosted the first stars.
Moreover, given the well-understood shape of the VAOs, they will allow us to separate cosmology from astrophysics, illuminating our knowledge about the end of the cosmic dark ages.

\acknowledgements

We wish to thank Aaron Ewall-Wice, David Pinner, and Jonathan Pober for helpful discussions, as well as Yacine Ali-Ha\"imoud, Cora Dvorkin, Anastasia Fialkov, Bradley Greig, Marc Kamionkowski, Andrei Mesinger, and especially Ely Kovetz for comments on a previous version of this manuscript.
We are also thankful to the creators of {\tt CAMB}, {\tt CLASS}, {\tt 21cmFAST} and {\tt 21cmSense}, as well as the libraries on which they depend, for making their codes publicly available.
Some computations in this paper were run on the Odyssey cluster supported by the FAS Division of Science, Research Computing Group at Harvard University.
This work was funded by a Department of Energy (DOE) Grant No. DE-SC0019018.

\bibliography{21cmBAO}

\appendix

\section{Dependence on the X-ray Spectrum}
\label{sec:AppXray}

In the main text we have assumed that the spectrum of emission of X-ray sources is given by a power-law with index $\alpha_X=1$ and a cutoff below energies $E_0=0.2$ keV~\cite{Mesinger:2010ne,Greig:2015qca,Visbal:2012aw}.
Nonetheless, it is possible that the first abundant X-rays in our Universe were emitted by compact binary objects, which are expected to have a higher-energy cutoff in their spectrum, with $E_0\approx0.5$ keV or even higher~\cite{Fragos:2013bfa}.
In that case, the mean-free path of X-ray photons will be elongated, erasing part of the small-scale power in the VAOs.
We illustrated this in Fig.~\ref{fig:jointFcoll}, where we considered the $F_{\rm coll}(\vcb)$ function averaged over distances of $R=20$ Mpc, typical of X-rays at 0.5 keV.
While the small-scale power is erased, the majority of the velocity-induced fluctuations are on large scales, which are unaltered by the X-ray propagation.

To ensure that our results can be generalized to any X-ray spectrum, we have performed a simulation with and without velocities, considering a larger X-ray cutoff of $E_0=0.5$ keV, similar to that of Ref.~\cite{Pacucci:2014wwa}.
As shown in Ref.~\cite{Greig:2017jdj}, the X-ray spectral index $\alpha_X$ does not alter results significantly, so we do not change it, and neither do we alter the total soft-band ($E_0\leq 2$ keV) X-ray luminosity $L_X$, as we varied that parameter in the main text.
In general, the change from a soft to a hard X-ray spectrum can delay the onset of the EoH significantly, and produce deeper 21-cm absorption~\cite{Fialkov:2014kta}.
This is not the case for $E_0\leq0.5$ keV~\cite{Greig:2017jdj}, since very low cutoffs result in very local X-ray heating, whereas for $E_0\gtrsim 1$ keV this delay can be important~\cite{Fialkov:2014kta}.

\begin{figure}[hbtp!]
	\includegraphics[width=0.5\textwidth]{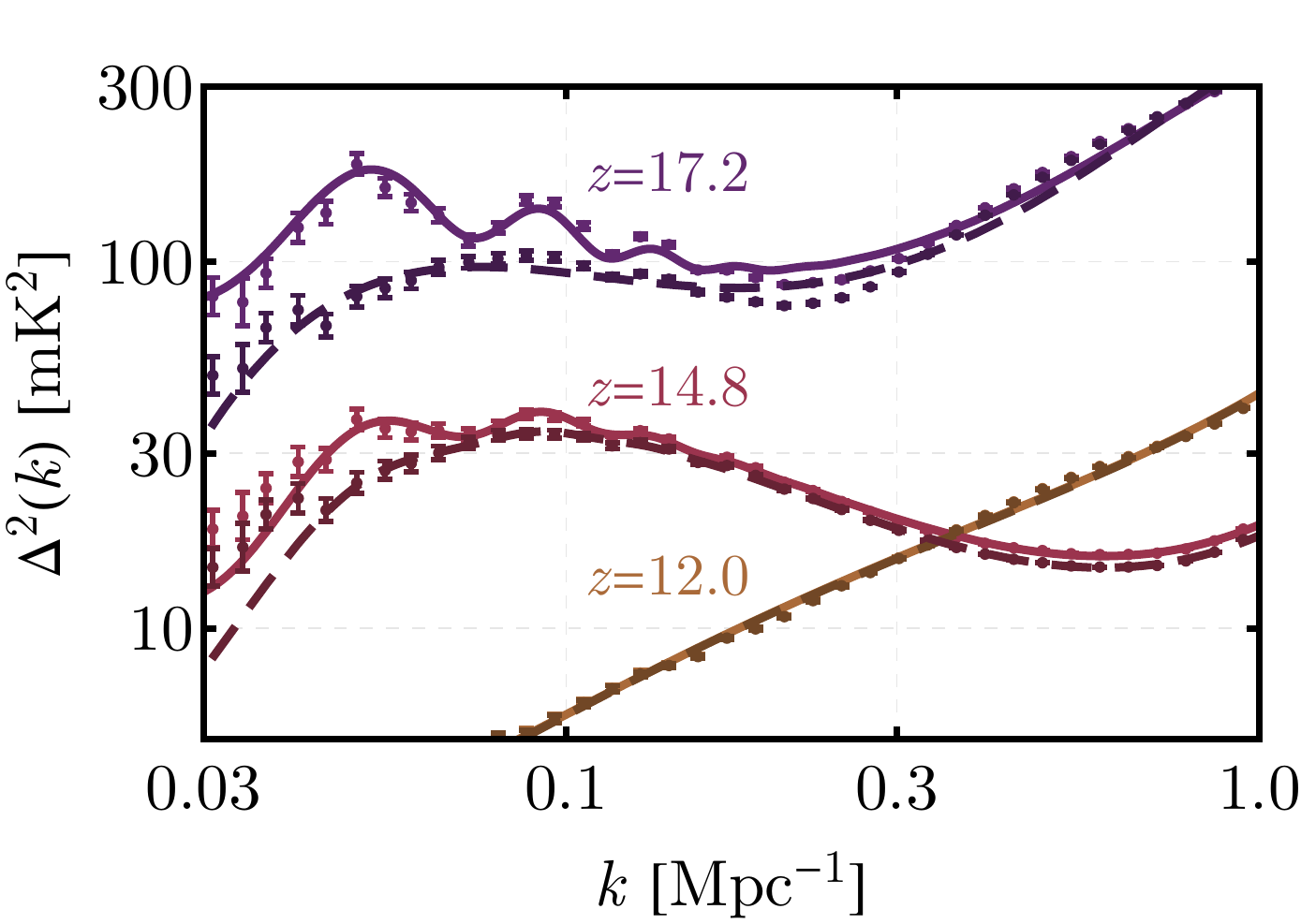}	
	\caption{21-cm power spectrum as a function of wavenumber $k$, where we have increased the minimum energy of the X-ray spectrum to $E_0=0.5$ keV.
	The redshifts correspond to the landmarks in the EoH: post-heating ($z=12.0$), at the transition to emission ($z_0=14.8$), and halfway through heating $(z=17.2)$.
	Solid lines are the fits to the simulations, using the modified window function for the new X-ray spectrum. Light colors indicate the inclusion of $\vcb$ fluctuations, whereas  darker colors and dashed lines have a fixed $\vcb=v_{\rm avg}$.
	}
	\label{fig:Pow21Xray05}
\end{figure}

We recalculated the X-ray window function with the new value of $E_0$ using Eq.~\eqref{eq:windowX}, confirming that the $\vcb$ fluctuations are quickly erased for $k\gtrsim 0.2$ Mpc$^{-1}$.
We fit for the results of the new simulation (with regular feedback), which we show in Fig.~\ref{fig:Pow21Xray05} at the three critical redshifts during the Epoch of Heating, both with $\vcb$ fluctuations and fixing $\vcb=v_{\rm avg}$ (to keep the same background).
First, we see that after heating is fully completed (at $z=12$) the velocities have a negligible effect on the 21-cm signal, as expected.
Then, both at the transition redshift (where $\VEV{T_{21}}=0$, corresponding to $z_0=14.8$) and the halfway point (defined as $\VEV{T_{21}} = 0.5\times\VEV{T_{21}}_{\rm min} $ and occuring at $z=17.2$) we see that the effect of velocities is as large as in our fiducial case , whereas the VAOs are suppressed at $k\gtrsim 0.2$ Mpc$^{-1}$ (cf.~Fig.~\ref{fig:21cmXray}).
This would hamper the detection of 21-cm VAOs, as smaller scales are typically freer of foregrounds, and therefore easier to measure.
Nonetheless, we see that even in this case there is a sizable effect at observable scales, which experiments like HERA could observe at a SNR of $\{ 3 ,7,  50 \}$ during the EoH under pessimistic, moderate, and optimistic foregrounds, respectively. This is slightly lower than in the  case studied in the text, due to the shorter mean-free path of X-ray photons.
As a test of the robustness of VAOs, we have attempted to model this case (with $E_0=0.5$ keV) with the window function of the case in the main text ($E_0=0.2$ keV).
While this produces a slightly worse fit (indicating that $E_0$ can potentially be constrained by the shape of VAOs, were they to be detected), the acoustic peaks do not change locations, as expected.

\section{Atomic-cooling Haloes}
\label{App:atomic}

Here we explore the effect of velocities in the 21-cm line assuming that Lyman-Werner feedback is fully saturated at all times, so that stars can only form through atomic cooling.
This requires haloes with a circular velocity roughly above $V_{\rm cool}=17$ km s$^{-1}$~\cite{Oh:2001ex,Barkana:2000fd,Bromm:2002hb} (translating into a minimum virial temperature of $T_{\rm cool}=10,500$ K), which have masses of $M \gtrsim 3\times 10^7\,M_\odot$ at $z\sim 20$.
For these haloes the effects of streaming velocities are not very significant, as they are formed of overdensities at scales larger than molecular-cooling haloes, and it is not expected that $V_{\rm cool}$ is directly modulated by $\vcb$.
Nonetheless, relative velocities can still leave a small imprint in the 21-cm line without molecular-cooling haloes, as we will explore here.

We perform a simulation under these assumptions, with the same fiducial parameters as in the main text. 
From this simulation, we showed in Fig.~\ref{fig:T21avg} the background evolution of $T_{21}$ when only atomic-cooling haloes are allowed, 
along with the cases studied in the main text.
There we see that at high redshifts the atomic-only case severely underpredicts the amount of stars formed, delaying the beginning of the LCE by $\Delta z \approx 3$, and the transition to the EoH by $\Delta z \approx 1$.
Nonetheless, as time evolves the LW feedback in our regular case becomes more important, so the atomic-only case transitions into emission only $\Delta z\approx0.3$ later than our regular-feedback case.
Moreover, by the end of our simulations ($z=12$) the regular-feedback case and the atomic-only are almost identical, as the LW feedback is nearly saturated, impeding star formation in molecular-cooling haloes. 
This showcases the need to properly include the redshift evolution of the LW feedback---plus relative velocities---to obtain the evolution of $T_{21}$ across the entire cosmic dawn.

\begin{figure}[hbtp]
	\includegraphics[width=0.48\textwidth]{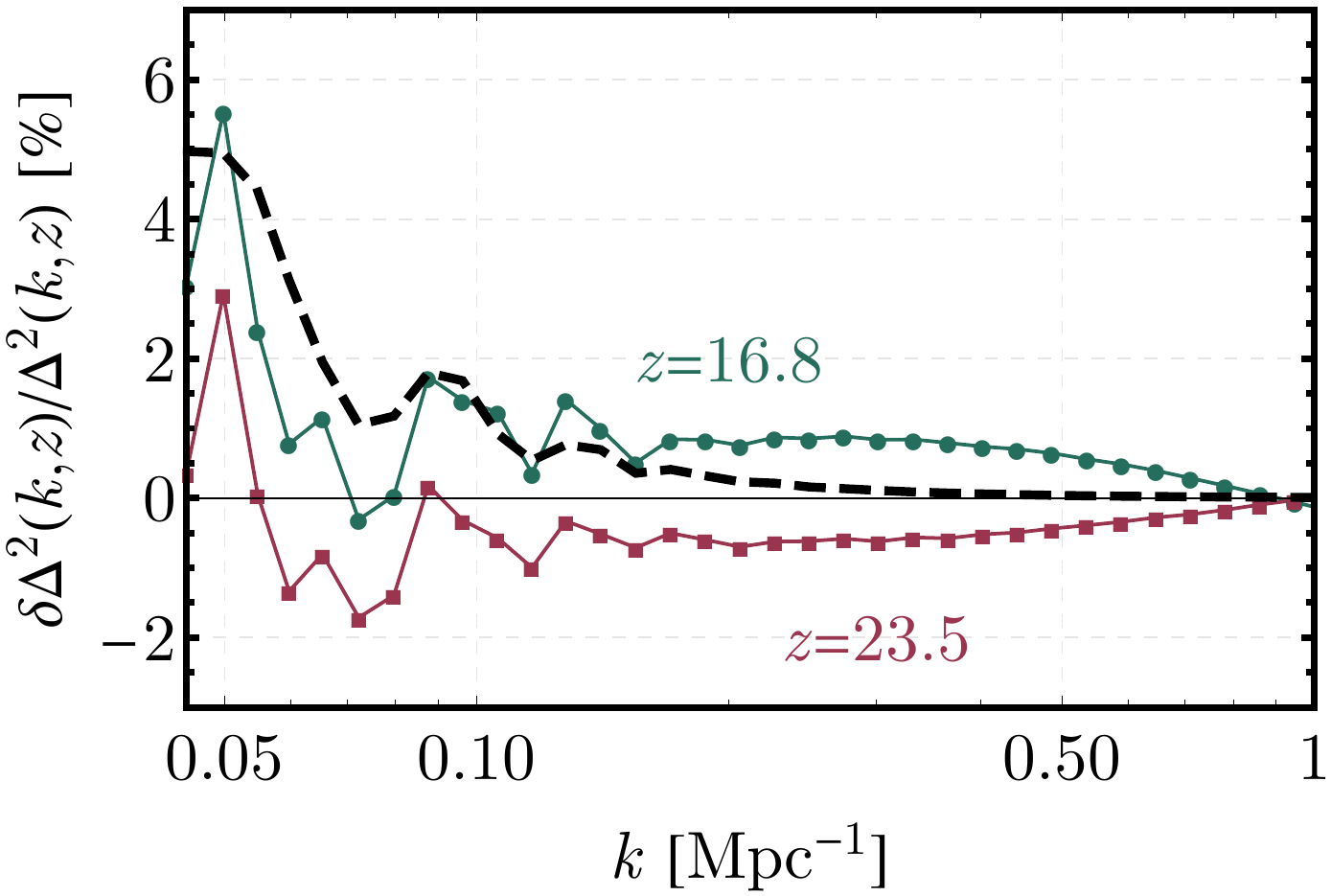}	
	\caption{Relative difference in the 21-cm power spectrum between the case with and without velocities, considering only haloes that can form stars through atomic cooling ($T_{\rm cool}=10^4$ K).
		We show results at $z=16.8$ and $23.5$, which correspond to the halfway points of the EoH and the LCE, where $\VEV{T_{21}} = 0.5\times \VEV{T_{21}}_{\rm min}$.
		The black-dashed line follows the shape of $\Delta^2_{v^2}(k)$ divided by the fiducial 21-cm power spectrum, shown for comparison.
	}
	\label{fig:atomicP21}
\end{figure}

Focusing on the atomic-only case for now, we show in Fig.~\ref{fig:atomicP21} the (relative) difference between the 21-cm power spectrum with and without relative velocities at $z=16.5$ and $23.5$, corresponding to the mid points of the EoH and the LCE in this case, where we see that even for these heavy haloes there is a small VAO signal, at the few-percent level.
In Fig.~\ref{fig:atomicP21} we have performed simulations with the same initial conditions in the cases with and without relative velocities, so the ``cosmic variance" (Poisson noise) mostly cancels when subtracting them~\cite{Seljak:2008xr}, leaving a cleaner VAO signal.
While the large-scale differences are due to the VAO, the small scale changes are due to the different backgrounds, and thus can potentially be mimicked by a redshift-dependent feedback.

\begin{figure}[tbp!]
	\includegraphics[width=0.48\textwidth]{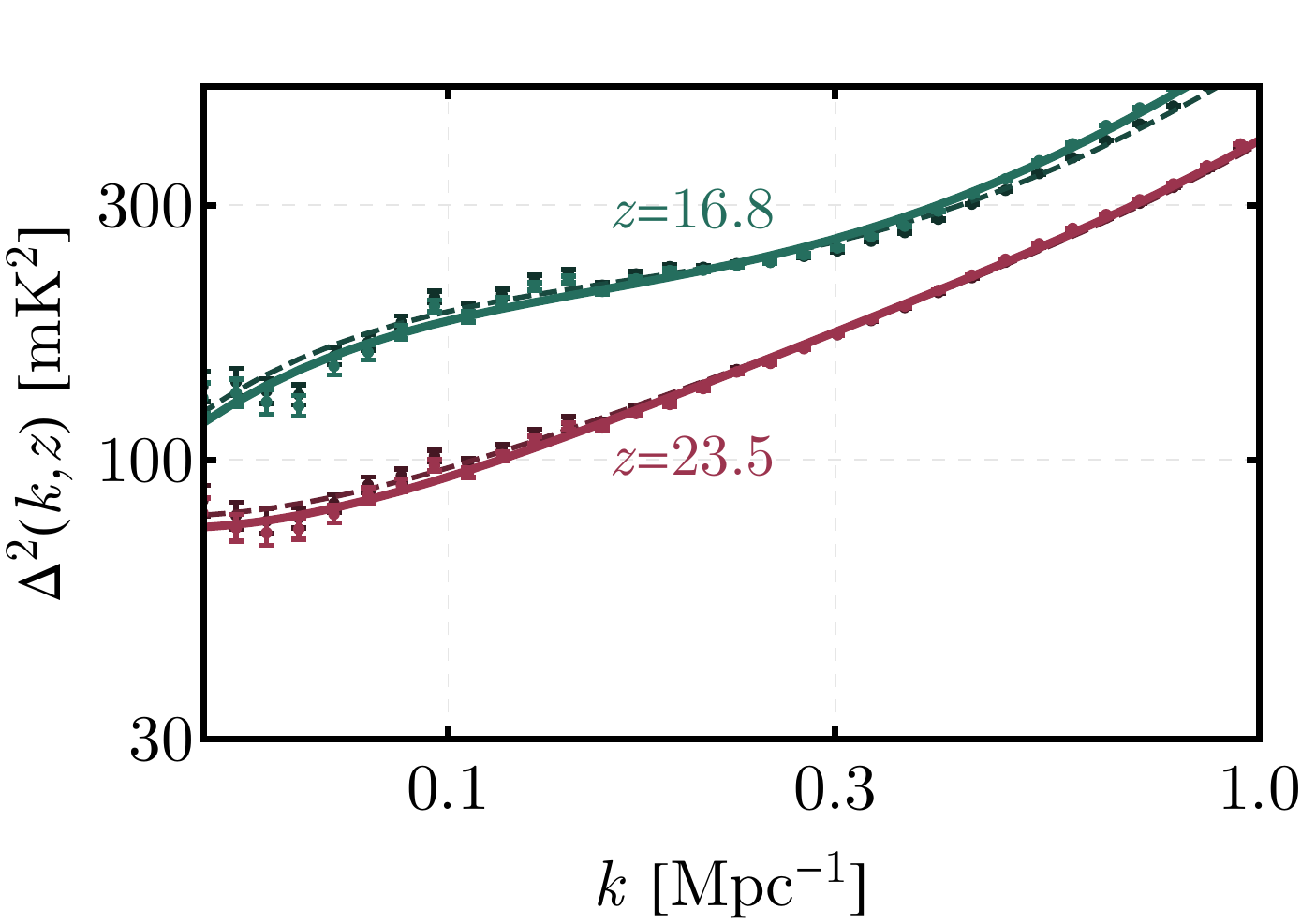}	
	\caption{21-cm power spectrum for standard {\tt 21CMMC} (in dark dashed lines) and {\tt 21cmvFAST} with $\vcb=0$ (in solid lines), assuming atomic cooling only.
		The minor differences arise from the slightly different background evolution in the two cases.
	}
	\label{fig:atomiccompare}
\end{figure}

We also compare our atomic-cooling case with the standard {\tt 21CMMC} output in Fig.~\ref{fig:atomiccompare}, as a cross check of our modifications to the code.
Here we have set $\vcb=0$, changed our transfer functions to match the fit from Ref.~\cite{Eisenstein:1997ik}, and switched our halo mass function to that of Ref.~\cite{Sheth:1999su}, as those are the settings in {\tt 21CMMC}~\cite{Greig:2015qca,Mesinger:2010ne,Greig:2017jdj,Greig:2018hja}.
With these modifications we see that our prescription very well matches the standard {\tt 21CMMC} output, as clear from Fig.~\ref{fig:atomiccompare}, where both simulations share the same initial conditions.
Nonetheless, in our prescription we compute $\delta_m(k,z)$ (and thus $F_{\rm coll}$) by evolving Eqs.~\eqref{eq:ODEs}, whereas in {\tt 21CMMC} a scale-independent growth factor is assumed. This produces small background differences in $\VEV{T_{21}}$ (of $\approx$ 1 mK), which induce a slight difference in the power spectra in Fig.~\ref{fig:atomiccompare} as well.

\section{Power Spectrum of Functions of $\vcb$}
\label{sec:Appvsq}

In the main text we argued that the power spectrum of any function of the relative velocity is, to a good approximation, proportional to that of the quantity
\be
\delta_{v^2} \equiv \sqrt{\dfrac{3}{2}} \left( \dfrac{v^2}{v_{\rm rms}^2} - 1\right),
\ee
as first shown in Ref.~\cite{Dalal:2010yt}.
Here we further strengthen their argument by directly calculating the power spectrum of a complete basis of functions, which can in principle reconstruct the power spectrum of any function of $v$.

We can write any function as
\be
f(v) = \VEV{f(v)} + \sum_n a_n \delta_v^{(n)},
\ee
where $a_n$ are some expansion coefficients, and we have defined
\be
\delta_v^{(n)} = \dfrac{\vcb^n}{\VEV{\vcb^n}} - 1,
\ee
so that $\VEV{\delta_v^{(n)}} = 0$ for all $n$.
Analytic functions of $\mathbf \vcb$ should only depend on even powers of $\vcb$, although we will also consider odd values of $n$  here.
The power spectrum of $f$ due to velocities will be given by
\be
\VEV{f(\mathbf k)f(\mathbf k')} = \sum_{n,m} a_n a_m \VEV{\delta_v^{(n)}(\mathbf k)\delta_v^{(m)}(\mathbf k')},
\ee
so in principle it is sufficient to calculate all cross spectra
\be
\VEV{\delta_v^{(n)}(\mathbf k)\delta_v^{(m)}(\mathbf k')} = (2\pi)^3 \delta_D(\mathbf k + \mathbf k') \Delta_{n,m}^2 (k),
\label{eq:Deltanm}
\ee
to find the power spectrum of any arbitrary function $f$ of the velocities.
It is this quantity $\Delta_{n,m}^2 (k)$ that we will focus on here.

\begin{figure}[hbtp]
	\includegraphics[width=0.48\textwidth]{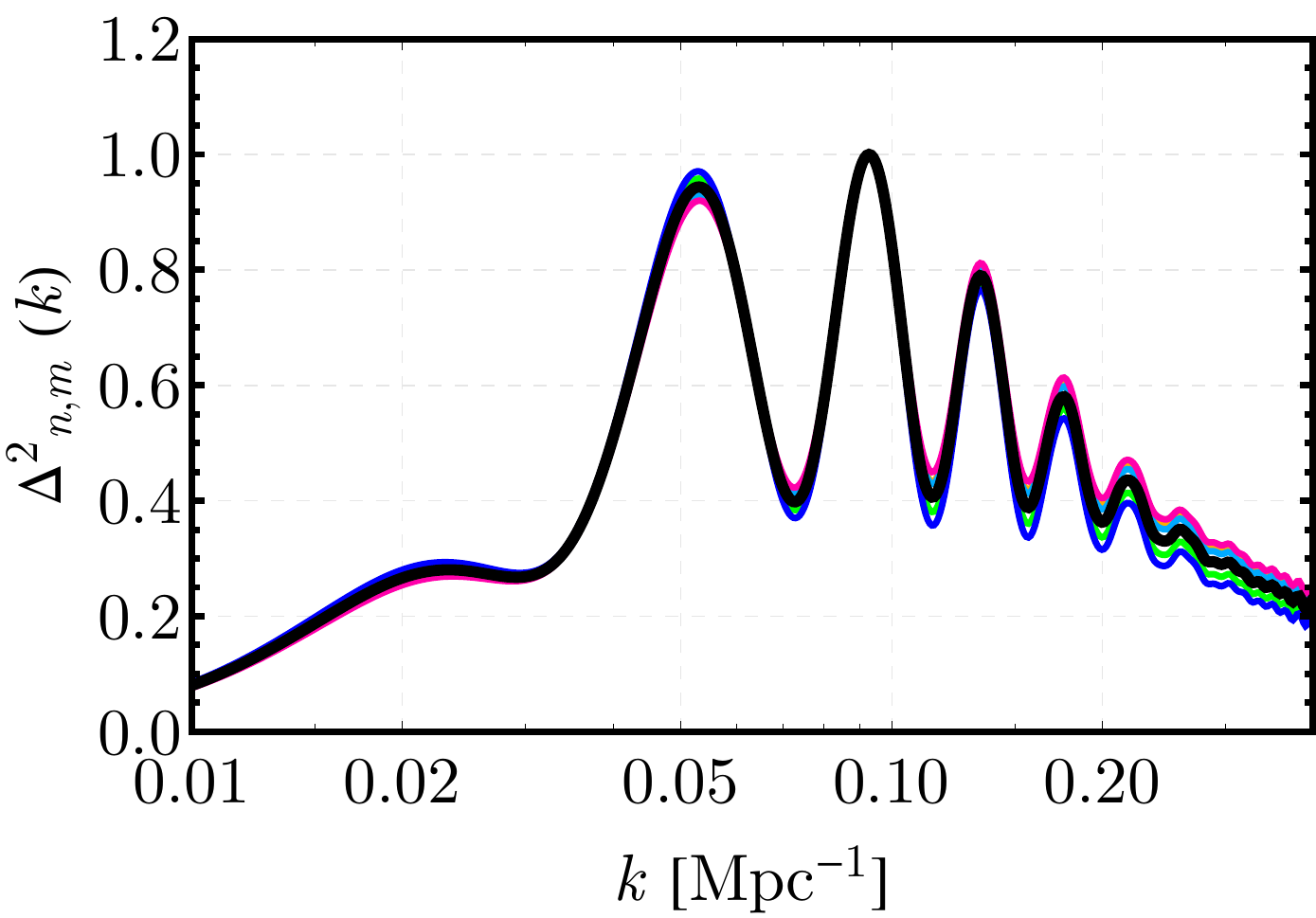}	
	\caption{Cross spectra of $\delta_v^{(n)}$ and $\delta_v^{(m)}$, as defined in Eq.~\eqref{eq:Deltanm}, normalized at $k=0.1\,\rm Mpc^{-1}$, as a function of wavenumber $k$.
	Different colors correspond to different values of $n$ and $m$ from 1 to 4, where the black thick line is the baseline $n=m=2$ case.
	This shows that the power spectrum of any generic function $f(\vcb)$ of the DM-baryon relative velocity has the same VAO shape.
	}
	\label{fig:Deltanm}
\end{figure}

We compute, following the approach in Refs.~\cite{Dalal:2010yt,Munoz:2018jwq,Ali-Haimoud:2013hpa}, the cross spectrum $\Delta_{n,m}^2 (k)$ of velocities, and show it in Fig.~\ref{fig:Deltanm}.
From this figure it is apparent that all the curves are very similar, with acoustic peaks essentially at the same locations and small differences in tilts.
Therefore, the power spectrum of any smoothly varying function of velocities can be well approximated by a linear combination of these very-similar curves.
We have found that the size of the power spectrum of any function of $v$ is proportional to $\left. d f(v)/dv\right|_{v=v_{\rm rms}}$, and only when setting this first derivative to zero we can find deviations from the $\Delta^2_{v^2}$ pattern that we have studied in this work.
We conclude, therefore, that it is unlikely that the specific dependence of $T_{21}$ on $\vcb$ can change the shape of the power spectrum of any function, whereas photon propagation throughout the Universe can have a larger impact.

\end{document}